\documentclass[aps,prb,article,superscriptaddress,twocolumn]{revtex4}
\usepackage[ansinew]{inputenc}
\usepackage{graphicx}
\usepackage{amsmath}
\usepackage{amsthm}
\usepackage{bm}
\usepackage{layout}
\usepackage{float}
\usepackage{txfonts}
\usepackage{amsfonts}
\usepackage{amssymb}%
\usepackage[ansinew]{inputenc}
\usepackage{graphicx}
\usepackage{amsmath}
\usepackage{amsthm}
\usepackage{bm}
\usepackage{layout}
\usepackage{float}
\usepackage{amsfonts}
\usepackage{amssymb}
\usepackage{appendix}
\usepackage{color}

\newsavebox{\smlmat}% Box to store smallmatrix content
\savebox{\smlmat}{$\left(
                     \begin{smallmatrix}
                       \mathrm{Re}U & -\mathrm{Im}U \\
                       \mathrm{Im}U & \mathrm{Re}U \\
                     \end{smallmatrix}
                   \right)
$}

\def\bb{\mathbf{b}}
\def\ba{\mathbf{a}}

\def\bx{\mathbf{x}}
\def\by{\mathbf{y}}
\def\bz{\mathbf{z}}
\def\be{\mathbf{e}}

\def\TT{\mathrm{T}}
\def\Tr{\mathrm{Tr}}
\def\bB{\mathbf{B}}

\def\ket#1{|#1\rangle}
\def\bra#1{\langle#1|}
\def\braket#1#2{\langle#1|#2\rangle}

\newtheorem{Lemma}{{\bf Lemma}}

\usepackage{bm}
\usepackage{graphicx}

\begin{document}

\title{The classical limit of a physical theory and the dimensionality of space}

\author{Borivoje Daki\'c}
\affiliation{Vienna Center for Quantum Science and Technology (VCQ), Faculty of Physics, University of Vienna,
Boltzmanngasse 5, A-1090 Vienna, Austria}
\affiliation{Centre for Quantum Technologies, National University of Singapore, 3 Science Drive 2, Singapore 117543}
\author{{\v C}aslav Brukner}
\affiliation{Vienna Center for Quantum Science and Technology (VCQ), Faculty of Physics, University of Vienna,
Boltzmanngasse 5, A-1090 Vienna, Austria} \affiliation{Institute of
Quantum Optics and Quantum Information (IQOQI), Austrian Academy of
Sciences, Boltzmanngasse 3, A-1090 Vienna, Austria}

\begin{abstract}

In the operational approach to general probabilistic theories one distinguishes two spaces, the state space of the ``elementary systems'' and the physical space in which ``laboratory devices'' are embedded. Each of those spaces has its own dimension-- the minimal number of real parameters (coordinates) needed to specify the state of system or a point within the physical space. Within an operational framework to a physical theory, the two dimensions coincide in a natural way under the following ``closeness'' requirement: the dynamics of a single elementary system can be generated by the invariant interaction between the system and the ``macroscopic transformation device'' that itself is described from within the theory in the macroscopic (classical) limit. Quantum mechanics fulfils this requirement since an arbitrary unitary transformation of an elementary system (spin-$1/2$  or qubit) can be generated by the pairwise invariant interaction between the spin and the constituents of a large coherent state (``classical magnetic field''). Both the spin state space and the ``classical field'' are then embedded in the Euclidean three-dimensional space. Can we have a general probabilistic theory, other than quantum theory, in which the elementary system (``generalized spin'') and the ``classical fields'' generating its dynamics are embedded in a higher-dimensional physical space? We show that as long as the interaction is pairwise, this is impossible, and quantum mechanics and the three-dimensional space remain the only solution. However, having multi-particle interactions and a generalized notion of  ``classical field'' may open up such a possibility.

\end{abstract}

\maketitle

\section{Introduction}

``Physical space is not a space of states'' writes Bengtsson in his article entitled ``Why is space three dimensional?''~\cite{Bengtsson}. Indeed, although the state space dimension for a macroscopic object is exponentially large (in the number of object's constituents), we still find ourselves organizing data into a three-dimensional manifold called ``space''. Why is this discrepancy? Can there be more dimensions? In past different approaches have been taken to show that the three-dimensional space is special, such as bio-topological argument~\cite{Freeman}, stability of planet orbits~\cite{Freeman}, stability of atoms~\cite{Ehrenfest} or elementary particle properties~\cite{Herbut}. The existence of extra dimensions has been proposed as a possibility for physics beyond the standard model~\cite{Kaluza,Klein,Antoniadis,Agashe,Dvali,Randal}.

In this work we will address the questions given above within the operational approach to general probabilistic theories~\cite{Barrett,Barnum06,Hardy}. There the basic ingredients of the theory are primitive laboratory procedures by which physical systems are prepared, transformed and measured by laboratory devices, but the systems are {\em not} necessarily described by quantum theory. General probabilistic theories are shown to share many features that one previously have expected to be uniquely quantum, such as probabilistic predictions for individual outcomes, the impossibility of copying unknown states (no cloning)~\cite{Barnum07}, or violation of Bell's inequalities~\cite{Bell,PRBox}. Why then nature obeys quantum mechanics rather than other probabilistic theory? Recently, there have been several approaches, answering this question by reconstructing quantum theory from a plausible set of axioms that demarcate phenomena that are exclusively quantum from those that are common to more general probabilistic theories~\cite{Fivel,Hardy,Hardy1,Fuchs,CBH,Grinbaum1,BruknerZeilinger,BruknerZeilinger1,Dariano,Grinbaum2,Goyal1,DakicBrukner,MassanesMueler,Dariano1,Massanes,Rau}.

In probabilistic theories the macroscopic laboratory devices are standardly assumed to be classically describable, but are not further analyzed. The ``position'' of the switch at the transformation device or the record on the observation screen have only an abstract meaning and are not linked to the concepts of position, time, direction, or energy of ``traditional'' physics (or to use Barnum's words ``the full, meaty physical theory'' is still missing~\cite{Barnum08}). As a result of the reconstructions of quantum theory, one derives a finite-dimensional, or countable infinite-dimensional, Hilbert space as an operationally testable, abstract formalism concerned with predictions of frequency counts in future experiments with no appointment of concrete physical labels to physical states or measurement outcomes. In standard textbook approach to quantum mechanics this appointment is ``inherited'' from classical mechanics and is formalized through the first quantization -- the set of explicit rules that relate classical phase variables with quantum-mechanical operators. However, these rules lack an immediate operational justification. This calls for a ``completion'' of operational approaches to quantum mechanics with the ``meaty physics''. Our work can be understood as a step in this direction.

In an operational approach one interprets parameters that describe physical states, transformations, and measurements, as the parameters that specify the configurations of macroscopic instruments in physical space by which the state is prepared, transformed, and measured. Within this approach it is natural to assume the state space and the physical space to be isomorphic to each other. The isomorphism of the two spaces is realized in quantum mechanics for the elementary directional degree of freedom (spin-1/2). The state space of the spin is a three-dimensional unit ball (the Bloch ball) and its dimension and the symmetry coincide with those of the Euclidian (non-relativistic) three-dimensional space in which classical macroscopic instruments are embedded. This was first pointed out by von Weizs\"{a}cker who writes~\cite{Weizsaecker}: ``It [quantum theory of the simple alternative] contains a two-dimensional complex vector space with a unitary metric, a two-dimensional Hilbert space. This theory has a group of transformations which is surprisingly near-isomorphic with a group of rotations in the real three-dimensional Euclidian space. This has been known for a very long time. I propose to take this isomorphism seriously as being the real reason why ordinary space is three-dimensional.'' In a different vein, Penrose demonstrated that the angles of three-dimensional space can be modeled by spin networks in semiclassical states of ``large spins''~\cite{Penrose} and Wootters showed a relation between the statistical distinguishability in quantum mechanics and geometry~\cite{Wootters}.

Whereas von Weizs\"{a}cker based his proposal on a mathematical isomorphism between the two spaces, there are very compelling physical evidences that they indeed are related to each other. The Einstein-de Haas effect~\cite{de Haas} as well as the Barnett effect~\cite{Barnett} demonstrate a deep relationship between magnetism, angular momentum, and elementary quantum spin. In the Einstein-de Hass effect an external magnetic field, generated by electric current through the coil surrounding a ferromagnet, leads to the mechanical rotation of the ferromagnet (or reversely, in the Barnett effect, a spinning ferromagnet can change its magnetization). The two effects phenomenologically demonstrate that the quantum spin is indeed of the same nature as the angular momentum of macroscopic rotating bodies as perceived in classical mechanics. One can therefore associate mathematical properties to the elementary quantum spin that are typical for a vector (more precisely, pseudo-vector) in a three-dimensional  space, such as three coordinates, orientation in space, or building the cross products with other vectors. For example, the precession of the spin in the external magnetic field (the Larmor precession) is due to torque on the spin, which is given by the cross product between the spin and the field.

If one assumes that quantum theory is universal\cite{Peres}, one should be able to arrive at an explanation of macroscopic devices (such as those for preparation, transformation and measurement of elementary spins) in terms of classical physics and three-dimensional space from within quantum theory. This would allow to invert the logic from the previous paragraph and argue that the symmetry of the classical angular momenta as embedded in the three-dimensional Euclidian space should follow from the symmetry of the elementary quantum spin. One could offer such an explanation
in the ``classical'' or ``macroscopic'' limit of quantum theory. It is known that the spin coherent states~\cite{Atkins,Radcliffe} -- which are the states of a large number of identically prepared elementary spins -- acquire an effective description of a classical spin embedded in the ordinary three-dimensional space under the restriction of coarse-grained measurements~\cite{Kofler}. These (macroscopic) states are ``robust'': they are stable with respect to small perturbations, such as those caused by repeated observations, giving rise to ``objective'' properties in the classical limit. For example, if one flips only a few spins of a ferromagnet, the system will turn into an orthogonal state, but we will identify it as the very same magnet at the macroscopic level. The macroscopic distinguishability can be reached only if a sufficiently large number of spins (of the order of square-root of the total number of spins) are flipped in which case we perceive it as a new state of magnetization.

The spin coherent states can serve as ``reference states'' with respect to which one can define the notion of ``direction''. Preparation, rotation or measurement of the elementary quantum spin along some ``direction'' has then only relative meaning with respect to such quantum reference frames~\cite{Rudolph,Dickson,Aharonov,Poulin,Poulin1} which become classical ones in the limit of a large number of spins constituting the coherent state. In the limit the spin coherent states can be understood as representing the classical magnetic field in which other quantum spins may evolve. Importantly, the group of transformations of an individual quantum spin is then generated by a rotationally invariant interaction between the spin and the coherent state, i.e. by a pairwise invariant interaction between the spin and each of the constituting spins of the coherent state~\cite{Poulin}. The invariance is required as there is no external reference frame. The spin coherent states define directions in terms of two (polar) angles in the three-dimensional Euclidian space, and thus give rise, through the relative angle, to the notion of ``neighboring'' orientations, without having such a notion from the very beginning.

We have seen that there are phenomenological and mathematical evidences for the isomorphism between the state space of elementary quantum spin and the physical space. Central for the argument are coherent states, which can be understood as representing macroscopic fields in the physical space, on one hand, and are class of the states in Hilbert space for which all the spins are prepared in the same quantum state, on the other hand.

The notion of coherent states is not exclusive for quantum theory but can be straightforwardly extended to general probabilistic theories as well, as a state of the collection of a large number of equally prepared elementary systems. It is legitimate to think that starting with the theory that differs from quantum theory and going into the limit of states with a large number of elementary systems and coarse-grained measurements one might arrive at ``classical physics'' embedded in a space of dimensions different than the one of our everyday life~\cite{BruknerPH}. For example, quaternionic quantum theory describing non-relativistic spin requires the physical space to have five dimensions, and the octavic quantum theory requires the space of nine dimensions~\cite{Brody}.

Here we investigate the possibility of having higher-dimensional physical spaces in the macroscopic (``classical'') limit. Our analysis is restricted to non-relativistic geometry of space (not space-time and not curved spaces) and directional degrees of freedom (spin). It is clear that one can imagine a vast variety of manifolds as possible candidates for the space (for example, as odd as the donut shape). Our focus here is onto the most natural generalization of the experienced (three-dimensional) non-relativistic space: the Euclidean $d$-dimensional isotropic space. We have seen that the symmetry of the state space of the elementary quantum spin (three-dimensional Bloch sphere) has the symmetry of the three-dimensional physical space. This strongly suggests that one needs to go outside of quantum framework to explore possibilities of higher-dimensional physical spaces. The natural choice to start with are systems for which the state space is $d$-dimensional Bloch sphere and we call them ``generalized spins''. They can be derived from an information-theoretic analysis~\cite{Paterek} and come as the most natural generalization of quantum spin. All such systems share fundamental features with the quantum spin, such as quadratic uncertainty relations for mutually unbiased (complementarity) observables, isotropic set of states, its rotational symmetry etc. They only differ in the dimension $d$ of the state space~\cite{footnote3}.

A large number of equally prepared generalized spins define a (generalized) spin coherent state. Under the restriction of coarse-grained measurements such spin coherent state acquires an effective description of a classical vector embedded in the $d$-dimensional space. One might think that the analogue with quantum theory can be developed further in that a spin coherent state can define the ``field of the magnet'' in which the elementary generalized spin can evolve, analogous to the Larmor precession but in a higher-dimensional physical space. With no preferred direction one would require the pairwise interaction between the generalized spin and each of the constituting spins of the coherent state to be invariant under the simultaneous group action on both (the rotational invariance). Here we show that no such interaction between the spin and the macroscopic field can generate the group of transformations of the spin unless its  state space and the physical space in which the field acts are both three-dimensional --  as in quantum theory and in our three-dimensional world.

In more precise terms we impose the following requirements on theory:
\begin{itemize}
  \item \emph{{\bf(Closeness)} The dynamics of the elementary system of the theory can always be generated through the invariant interaction of the system with the macroscopic device that itself is obtained from within theory in the macroscopic limit.}
  \item \emph{{\bf(Macroscopic states)} The macroscopic transformation device (``magnetic field'') is in a coherent state in which the constituting elementary systems are all equally prepared.}
\end{itemize}
The two requirements can be fulfilled only if {\bf \em the symmetry of the elementary system and of the macroscopic device by which the system is transformed are both those of the Euclidean $d$-dimensional space}. If the elementary interactions between the elementary systems are {\em pairwise} the underlying theory is quantum theory and $d=3$.

An important restriction under which our result is obtained is that the generalized spin interacts {\em pairwise} with each single spin constituting the large spin in the coherent state. We show that if we relax this assumption, there are group invariant interactions between three or more generalized spins. This means that the spin under consideration could interact with several other spins, each one belonging to a different coherent state, and that such interaction could generate the group of transformation of the spin. The notion of the ``field of the magnet'' would then be extended such that it is represented not by a single but several coherent states. This opens up a possibility of having higher-dimensional Euclidian physical spaces compatible with underlying generalized probabilistic theory different from quantum theory. Nonetheless, we leave the question open of whether such a theory can be fully constructed in a mathematically consistent way.

In a recent work~\cite{MS}, M\"uller and Masanes gave an information-theoretic analysis of the relationship between the geometry of the state space of an elementary system (directional degree of freedom) and the classical space in which the macroscopic devices are embedded. In their work, they consider ``spin'' as an elementary directional degree of freedom to be measured by a macroscopic measurement device (``generalized Stern-Gerlach magnet'') that can be oriented along arbitrary direction in $d$-dimensional physical space. Assuming that any spatial direction can be encoded in a physical state of the spin and no further
information is encoded in the state, they derive that the state space is the $d$-dimensional Bloch sphere. In the next step, they show that such systems can exhibit continuous non-trivial dynamics only in three dimensions with the constraint to the {\em locally-tomographic} theories~\cite{Araki,Bergia,WoottersLT,Mermin}. 

In the present approach, we take a different route. From the very beginning we consider the systems that have $d$-dimensional Bloch sphere as the state space and obtain the dimensionality of the physical space by the requirement that the theory is ``closed''. In a probabilistic theory the dynamics of a single system is assumed to be generated by an external field of macroscopic devices. In a closed probabilistic theory the fields are not notions from ``outside'' of the theory, but are obtained from within it in the macroscopic limit. Furthermore, we extend the study to the more general class of theories that are not in general {\em locally tomographical}~\cite{Gullio}. The assumption of so called {\em local-tomography} states that the global state of a composite system can be learned trough local statistics. We allow for more general situations where the state of a composite system may include a set of {\em global parameters} that cannot be learned trough local statistics but trough the global (entangling) measurements on a whole system~\cite{HardyWotters}. The prototype of the theory that involves global parameters is quantum mechanics based on real amplitudes~\cite{Stueckelberg}. This theory can be reconstructed within an information-theoretic approach~\cite{HardyWotters}. Most of the previous information-theoretic reconstructions of quantum theory~\cite{Hardy,DakicBrukner,Massanes,Dariano1}, as well as the work of Ref.~\cite{MS} adopt local tomography (e.g. directly eliminating real quantum mechanics), in contrast to our work here.

In conclusion, we reconstruct, the three-dimensional space and quantum mechanics trough the macroscopic limit under the constraint of pairwise elementary interactions. Interestingly, higher-dimensional space may arise in the limit if one allows for multipartite elementary interactions, i.e. ternary and more.

\section{The classical limit of quantum theory and three-dimensionality of space}

In the operational approach to quantum mechanics the notion of quantum state refers to a well-defined configuration of the macroscopic instrument by which preparation of the state is defined. For example, the  ``horizontal'' polarization of photon is specified by the ``reference direction'' of a classical object relative by which the polarization is prepared, such as the plane of the polarizing filter. On the other hand, if quantum mechanical laws are universal, then macroscopic, classical objects, such as polarizing filters, themselves should allow a description from within quantum mechanics.

One can consider a macroscopic object as a collection of large number of elementary quantum systems, which are in one of ``macroscopically distinct states''. The latter are defined as quantum states that can still be differentiated even if the measurement precision is poor and one performs coarse-grained measurements. The states can be repeatedly measured by different observers or copied with negligible disturbance. They are ``robust'' under disturbance or losses of a sufficiently small number of constituent quantum systems. These properties give rise to a level of ``objectivity'' of the macroscopically distinct states among the observers. A good example of such ``classical'' states are large spin-coherent states~\cite{Atkins,Radcliffe} under the restriction of coarse-grained measurements. For the spin-$J$ system the spin coherent states are defined as the eigenstates with the largest eigenvalue of spin projection along direction $\vec{n}$ :
\begin{equation}
\hat{J}_{\vec{n}}\ket{\vec{n}}=J\ket{\vec{n}},
\end{equation}
where $\hat{J}_{\vec{n}}=\vec{n}\hat{\vec{J}}$ and $\vec{n}=\sin\theta\cos\phi\vec{e}_x+\sin\theta\sin\phi\vec{e}_y+\cos\theta\vec{e}_z$ (with no external reference frame assumed, $\vec{n}$ should be understood as a parametrization of the spin state with no further immediate physical interpretation). Their expansion in the eigenbasis of $\hat{J}_z$ reads:
\begin{equation}
\ket{\vec{n}}=\sum_{m=-J}^{J}{2J \choose J+m}^{\frac{1}{2}}(\cos\frac{\theta}{2})^{J+m}(\sin\frac{\theta}{2})^{J-m}e^{-im\phi}\ket{m}.
\end{equation}

The spin-$J$ particle can be considered as a composite system consisting of $N$ spin-1/2 particles. The spin-$J$ coherent state is then the product state of $N$ equally prepared spin-1/2 particles ($J=N/2$)
\begin{equation}
\ket{\vec{n}}=\ket{\vec{n}}_1\ket{\vec{n}}_2\dots\ket{\vec{n}}_N
\end{equation}
In the limit of large $J$ (or large $N$) the spin coherent states acquire the properties of ``classical'' states. The probability of obtaining outcome $m$ of $J_z$ is given by the binomial distribution $p_m=|\braket{m}{\vec{n}}|^2$. In the limit it reduces to the normal distribution:
\begin{equation}\label{normal distribution}
p_m=\frac{1}{\sqrt{2\pi}\sigma}e^{\frac{(m-\mu)^2}{2\sigma^2}},
\end{equation}
where $\sigma=\sqrt{N}\sin\theta$ is the width of distribution and $\mu=N/2\cos\theta$ is the mean value. The overlap between two spin-coherent states
\begin{equation}\label{spin-coh overlap}
|\braket{\vec{n}_1}{\vec{n}_2}|^2=\left(\frac{1+\vec{n}_1\vec{n}_2}{2}\right)^N\longrightarrow\delta_{\vec{n}_1,\vec{n}_2},
\end{equation}
becomes exponentially small in the limit of large $N$.

The uncertainty of measuring $J_z$ is given by the standard deviation $\sigma$. Under the restriction of coarse-grained measurements where the outcomes are merged into ``slots'' of size much larger than the standard deviation, the Gaussian cannot be distinguished anymore from the delta function~\cite{Kofler} and the spin-coherent states become effectively ``classical vectors'' in three-dimensional space.

There are two independent ways in which large spin-coherent states can be said to induce the properties of the physical space. Firstly, they can be used to define the ``reference direction'' in a three-dimensional space, though one lacks this notion in the abstract Hilbert space formulation of quantum theory to start with. With no external reference frame only rotationally invariant observables can be measured, such as the total spin length. Consider a ``large'' spin of length $J$ in a spin-coherent state $|\vec{n}\rangle$ and a ``small'' spin of length 1/2. It can be shown that the probability distribution for the outcomes $J+1/2$ (``aligned'') and $J-1/2$ (``anti-aligned'') of the total spin length approaches the probability distribution for the outcomes of spin projection of the spin-1/2 along the direction $\vec{n}$ in the classical limit ($N\rightarrow +\infty$)~\cite{Poulin,Rudolph}. In that way, the spin-coherent states define the complete {\em set of measurements} for the elementary spin. The set has the same dimensionality and the symmetry as the three-dimensional Euclidian physical space. We call these {\em static} properties of the space.

Secondly, spin-coherent states can generate non-trivial {\em dynamics} in three-dimensional space. A macroscopic spin in a coherent state can serve as an ``external magnetic field'' around which another spin can precess, i.e. it serves as a {\em transformation device} for the elementary spins~\cite{Poulin}. Since there is no preferred direction beside the one defined by the large spin one requires the interaction between the elementary spin and the large spin to be {\em rotationally invariant}. To illustrate it consider the situation as given in Figure~\ref{Spin Coherent Rotation} (left). A single spin-$1/2$ particle interacts with $N$ spins prepared in a coherent state along direction $\vec{n}$. Total interaction Hamiltonian is the sum of all pairwise interactions $H=\sum_{n=1}^NH^{(0n)}$ where $H^{(0n)}$ labels the interaction between the single spin and $n$th spin of the macroscopic system. There is only one rotationally invariant Hamiltonian, that is the Heisenberg spin-spin interaction $H^{(0n)}=J_n\vec{\sigma}_0\vec{\sigma}_n$, where $J_n$ is the coupling constant. It can be shown that in the macroscopic limit the elementary spin only negligibly affects the state of a large spin and the dynamics of the elementary spin becomes unitary~\cite{Poulin}:
\begin{equation}\label{unitary dynamics}
e^{itH}\ket{\psi}\ket{\vec{n}}\approx (e^{itH_{eff}}\ket{\psi})\ket{\vec{n}},
\end{equation}
where $H_{eff}=\vec{B}\vec{\sigma}$ is the effective Hamiltonian and  $\vec{B}$ represents the strength of ``macroscopic field'' around which the ``small spin'' precesses (see Appendix \ref{App_Spin_Dyn} for details).

\section{Generalized spins and higher-dimensional Space}

\subsection{Single System}\label{Single Gbit}

%%%%%%%%%%%%%%%%%%%%%%%%%%%%%%%%%%%%%%%%%%%%%%%%%%%%%%%%%%%%%%%%%%%%%%%%%%
\begin{figure}\centering
\includegraphics[width=9cm]{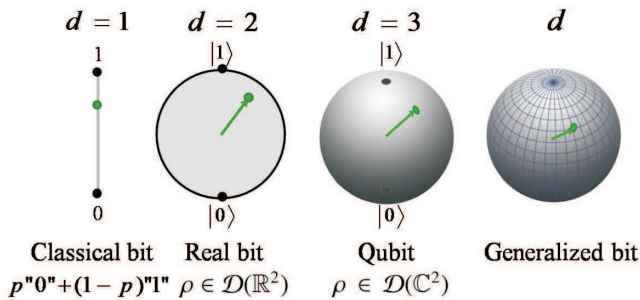}
\caption{Figure taken from Ref.~[\onlinecite{DakicBrukner}]. State spaces of a generalized spin or generalized bit (two-level system). The minimal number of real parameters $d$ is needed to specify the (mixed) state completely. From left to right: A classical bit with one parameter (the weight $p$ in
the mixture of two bit values), a real bit with two real parameters (state $\rho \in \mathcal{D}(\mathbb{R}^2)$ is represented by $2 \times 2$ real density matrix), a qubit (quantum bit) with three real parameters (state $\rho \in \mathcal{D}(\mathbb{C}^2)$ is represented by $2 \times 2$ complex density matrix) and a and a generalized bit for which $d$ real parameters are needed to specify the state. In the classical limit, a theory of elementary system with $d$ parameters gives rise to physics of macroscopic, classical ``fields'' embedded in $d$-dimensional physical space (see main text).}
\label{Gen Spin}
\end{figure}
%%%%%%%%%%%%%%%%%%%%%%%%%%%%%%%%%%%%%%%%%%%%%%%%%%%%%%%%%%%%%%%%%%%%%%%

Is there a microscopic theory that in its macroscopic limit leads to classical physics embedded in a physical space of dimension higher than three? Following previous discussions one can expect that the elementary (two-level) system with the $d$-dimensional sphere $\mathcal{S}^{(d-1)}$ as state space gives rise to coherent states and ``magnetic fields''  embedded in a $d$-dimensional Euclidean space in the macroscopic limit. Such an elementary system is non-quantum because it represents a two-level system with more than three degrees of freedom. Within the information-theoretic framework of generalized theories, such generalized bit (here called ``generalized spin'') is derived as the most natural generalization of qubit -- the system that is fundamentally limited to the content of one bit of information~\cite{Paterek,DakicBrukner}. Other information-theoretic approaches lead to the derivation of the same class of systems, e.g. by adopting {\em information causality}~\cite{Pawlowski} or {\em continuous reversible dynamics}~\cite{Massanes,MassanesMueler}.

The state of generalized spin is represented by a vector in a $d$-dimensional real space, $\bx=(x_1,\dots,x_d)$. The probability $ P_1(\bx,\by)$ to obtain the spin along direction $\by$ when the state is prepared along direction $\bx$ is expressed trough the generalized Born rule~\cite{DakicBrukner}:
\begin{equation}\label{Born1}
P_1(\bx,\by)=\frac{1}{2}(1+\bx^{\TT}\by).
\end{equation}
The set of pure states satisfy $P(\bx,\bx)=1$ and is represented by a unit sphere $\mathcal{S}^{d-1}$ in $d$-dimensions (see Figure~\ref{Gen Spin}). The characteristic feature differentiating between the theories is the number $d$ of parameters required to describe the state completely. For example, classical probability has one parameter, real quantum mechanics has two, complex (standard) quantum mechanics has three and the one based on quaternions has five parameters. A lower-order theory of the single system can always be embedded in a higher-order ones in the same way in which classical theory of a bit can be embedded in qubit theory.

Following the operational approach we assume that the continuous reversible transformations of macroscopic devices acting upon the system generates the continuous reversible transformation of the state of the system. Therefore, the set of physical transformations is a continuous (Lie) group. Furthermore, if an arbitrary reversible transformation of the states can be realized manipulating the macroscopic device, then the group of physical transformations is transitive on a sphere~\cite{Montgomery,Borel}, i.e. any pure state can be transformed to any other in a continuous fashion. We will consider minimal group transitive on $\mathcal{S}^{d-1}$, which is thus necessarily within the set of physical transformations (see Appendix~\ref{App_GT on Spheras}). The existence of such ``reversible transformations of macroscopic devices'' is usually assumed {\it ad hoc}. The aim of this work is exactly to show that they do no always exist, if the macroscopic devices are not considered ``outside'' of the theory, but are required to be obtained from within it in the classical limit.

\subsection{Generalized Spin-Coherent States}

Generalized spin-coherent states can be straightforwardly introduced in generalized probabilistic theories. For every dimension $d$, they are collections of $N$ equally prepared generalized spins. The preparation can be parameterized by a direction $\vec{n}$ in a $d$-dimensional space. Equations~\eqref{normal distribution} and~\eqref{spin-coh overlap}, derived in quantum theory, remain valid here as well. In the macroscopic limit of large $N$, the effective description of the coherent states is that of classical vectors embedded in a $d$-dimensional Euclidian space. We address here the question of whether generalized spin coherent states can generate non-trivial dynamics of individual spins in the space, similarly as the one given by equation~\eqref{unitary dynamics}. We will next show that with pairwise invariant interaction between elementary spins this is not possible except when $d=3$. We then discuss possible generalizations of our approach to multi-spin invariant interactions that might give rise to non-trivial dynamics in higher-dimensional spaces.

\section{Dynamics and Macroscopic Limit}\label{Section: dynamics and macro limit}

\subsection{The composite system}

In order to describe interactions between two or more generalized spins we need to introduce a representation of the composite system. One of the characteristics of both classical and quantum probabilistic theory is the local tomography~\cite{Araki,Bergia,WoottersLT,Mermin}, namely the property that the global state of a composite system is completely determined by the statistics of local measurements. For example, a state of two classical bits $\vec{p}=(p_{00},p_{01},p_{10},p_{11})$, where e.g. $p_{01}$ denotes the probability to obtain ``spin up'' on the first spin and ``spin down'' on the second one, can be equivalently represented by three numbers $(x,y,t)$:
\begin{eqnarray}
x&=&p_{00}+p_{01}-p_{10}-p_{11},\\
y&=&p_{00}-p_{01}+p_{10}-p_{11},\\
t&=&p_{00}-p_{01}-p_{10}+p_{11}.
\end{eqnarray}
The local statistics is given by mean values $x$ and $y$ of probabilities measured on the first and the second spin, respectively, whereas $t$ is the mean value of correlation (difference between the probabilities that the two spins are the same and that they are different). Similarly, the density matrix $\rho$ of two qubits can be decomposed as
\begin{equation}
\rho=\frac{1}{4}(\openone\otimes\openone+\sum_{i=1}^{3} x_i\sigma_i\otimes\openone+ \sum_{j=1}^{3} y_j\openone\otimes\sigma_j+\sum_{i,j=1}^{3} T_{ij}\sigma_i\otimes\sigma_j),
\end{equation}
where $\sigma_i,~i=1,2,3$, are Pauli operators. Vectors $\bx=(x_1,x_2,x_3)$ and $\by=(y_1,y_2,y_3)$ are called local Bloch vectors and are the mean values of the Pauli operators and $T$ is the $3\times3$ correlation matrix with elements $T_{ij}=\langle \sigma_i\sigma_j\rangle$.

Not all generalized probabilistic theories fulfill local tomography; an example is quantum mechanics based on real amplitudes. For the real bit only two Pauli matrices $\sigma_1$ and $\sigma_3$ correspond to physical observables, because $\sigma_2$ is a complex matrix. However, $\sigma_2\otimes\sigma_2$ is a real matrix, and thus it corresponds to a physical observable, although it cannot be measured locally. In general, a real density matrix $\rho$ can be represented in  a form
\begin{equation}
\rho=\frac{1}{4}(\openone\otimes\openone+\sum_{i=1}^{3}x_i\sigma_i\otimes\openone+\sum_{j=1}^{3}y_j\openone\otimes\sigma_j+\sum_{i,j=1}^{3}T_{ij}\sigma_i\otimes\sigma_j+\lambda\sigma_2\otimes\sigma_2),
\end{equation}
where $\lambda$ is a global parameter. Therefore, we can represent the state of a composite system by 4-tuple $(\bx,\by,T,\lambda)$.

We now introduce a representation of the composite system of two generalized spins. Firstly, we assume that local measurements on individual spins are well defined (i.e. probabilities for local measurements are non-negative and they sum up to one). Secondly, if the subsystems of a composite system are emitted from two independent sources, we assume that the joint probability distribution is factorizable. Consequently, one can define properly the set of product states as triples $(\bx,\by,T_p)$, where $T_p=\bx\by^{\TT}$. However, for general non-product states there might be some global parameters missing in the state description. Therefore, in the general case we associate a 4-tuple $\vec{\psi}_{12}=(\bx,\by,T,\Lambda)$ to the state of a composite system, where $\bx,\by$ are the local Bloch vectors, $T$ is a $d\times d$ real matrix that represents correlations and $\Lambda=(\lambda_1,\lambda_2,\dots)$ is a collection of global parameters that can be present in the state description but are not accessible trough statistics of local measurements.

We define the probability distribution for obtaining the two local local spins ``up'' along measurement directions $\ba,\bb$ to be
\begin{equation}\label{prob rule}
P_{12}(\vec{\psi}~|~\ba,\bb)=\frac{1}{4}(1+\bx\ba+\by\bb+\ba T\bb),
\end{equation}
where $\vec{\psi}=(\bx,\by,T,\Lambda)$ is the state of the composite system. The formula can also be interpreted as the overlap between the state $\vec{\psi}$ and product state $\vec{\phi}_p=(\ba,\bb,\ba\bb^{\TT},0)$:
\begin{equation}\label{Born12}
P_{12}(\vec{\psi},\vec{\phi}_p)=\frac{1}{4}(1+\vec{\psi}^{\TT}\vec{\phi}).
\end{equation}

A general state of $N$ spins is represented by $\vec{\psi}_N=(\bx_1,\dots,\bx_N,T_{12},\dots,T_{123},\dots,T_{1\dots N},\Lambda)$, where $\bx_i$ is the local Bloch vector of the $i$-th spin, tensors $T_{i_1i_2\dots}$ represents correlations (two-spin, three-spin etc.) and $\Lambda=(\Lambda_{12},\Lambda_{13},\dots,\Lambda_{123},\dots)$ is the set of all global parameters, where, for example, $\Lambda_{123}$ is the global parameter related to subsystems $1,2$ and $3$.

\subsection{Dynamics in Macroscopic Limit}

%%%%%%%%%%%%%%%%%%%%%%%%%%%%%%%%%%%%%%%%%%%%%%%%%%%%%%%%%%%%%%%%%%%%%%%%%%
\begin{figure*}\centering
\includegraphics[width=\textwidth]{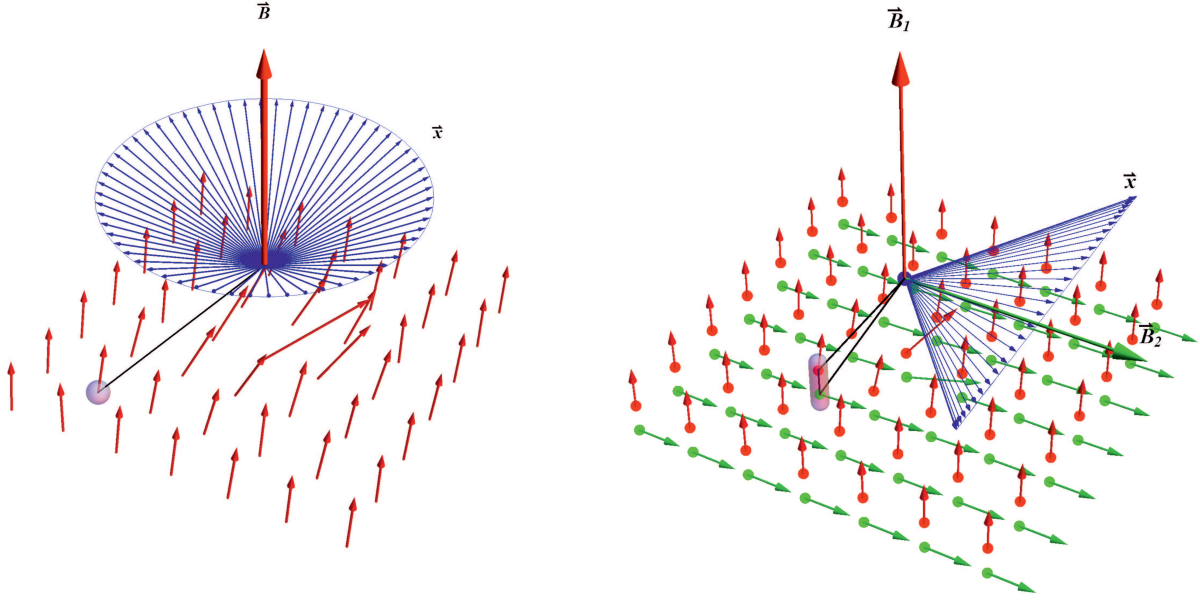}
\caption{Dynamics of the generalized spin as generated by its interaction with a single coherent state in three-dimensional space (left) or with a pair of coherent states in four-dimensional space (right). The coherent state is a collection of a large number of equally prepared constituent spins which are distributed here on a regular lattice. With no pre-existing reference direction all interactions are assumed to be rotationally invariant. In the macroscopic limit of an infinite large coherent states the effect of the spin on the coherent state is negligible and the dynamics becomes separable (i.e. the spin evolves according to the unitary evolution and the coherent state remains unchanged). (Left) Rotation of the quantum spin in three dimensions. The spin $\vec{x}$ interacts pairwise with each constituent spin of the coherent state $\vec{B}$. In the macroscopic limit this results in an effective precession of spin $\vec{x}$ around the classical macroscopic field generated by the spin coherent state $\vec{B}$. (Right) Rotation of the generalized spin in four dimensions. The spin $\vec{x}$ interacts via a three-particle interaction with each spin-pair, where one spin (red) of the pair belongs to coherent state $\vec{B_1}$ and the other one (green) to coherent state $\vec{B_2}$. In the macroscopic limit the effective dynamics of the generalized spin is rotation in the plane orthogonal to two macroscopic fields which are represented by the two coherent states $\vec{B}_1$ and $\vec{B}_2$. In the figure it is shown a projection of the dynamics in three dimensions. }
\label{Spin Coherent Rotation}
\end{figure*}
%%%%%%%%%%%%%%%%%%%%%%%%%%%%%%%%%%%%%%%%%%%%%%%%%%%%%%%%%%%%%%%%%%%%%%%

Dynamics of an individual generalized spin as generated by a transformation device is given by:
\begin{equation}\label{diff transform}
\frac{dx_i}{dt}=g_{ij}x_j,
\end{equation}
where $[G]_{ij}=g_{ij}$ is the generator of evolution and $t$ is the parameter of the transformation, usually taken to be time. (Here and in the rest of the article the summation over repeated indices is always assumed.) The integral version of the formula reads
\begin{equation}\label{transform}
\bx(t)=U(t)\bx(0),
\end{equation}
where $U(t)=\exp(tG)$ is the reversible transformation that belongs to the group of transformation $\mathcal{G}$ of the generalized spin. Our main objective is to investigate if such a dynamics can be obtained as a mean field approximation of the theory. (Note that in quantum mechanics this is the case and Eq. (\ref{transform}) is equivalent to Eq. \eqref{unitary dynamics}.) More precisely, we want to find out whether formula  (\ref{transform}) for the dynamical evolution of an individual generalized spin can be seen as a consequence of its interaction with a system composed of a large number of generalized spins (e.g. in coherent state). A negative answer to this question would indicate that the theory is not {\it closed}.

We represent a single spin by its local Bloch vector $\bx$ and the ``large'' system by a state $\vec{\psi}_N$.
In the limit of large $N$, the following holds
\begin{equation}\label{DYN}
W_N(t)\vec{\psi}_N\otimes\bx=\vec{\psi}_N\otimes U(t)\bx+\vec{O}(N,t),
\end{equation}
where $W_N$ represents the joint evolution of the system and of the field after duration $t$ of the interaction. The state $\vec{\psi}_N\otimes\bx$ represents the product state of a joint system (large $+$ small system), in a sense that all the correlation tensors are factorized. If the dynamics of the small spin can be reproduced from the interaction, one has $\vec{O}(N,t)\rightarrow0$ in the limit when the number of spins $N$ goes to infinity. Consequently, one recovers the equations \eqref{transform} and \eqref{diff transform} exactly, the initial state $\vec{\psi}_N$ of the large system remains almost unchanged, and the dynamics factorizes.

\subsection{Pairwise interaction}

Here we assume that all the interactions are pairwise at the elementary level. In section \ref{Going beyond d=3} we will relax this assumption.
The state of the composite system of two elementary generalized spins is represented by $\vec{\psi}_{12}=(\bx,\by,T,\Lambda)$. The dynamical law reads $\vec{\psi}_{12}(t)=W_{12}(t)\vec{\psi}_{12}(0)$, where $t$ is the duration of interaction. One has $W_{12}(t)=\exp(t H)$, where $H$ is the generator of the interaction $W_{12}$. The differential version of the dynamical law reads:
\begin{eqnarray}\label{gen inter1}
\frac{dx_i}{dt}&=&a_{ij}x_j+b_{ij}y_j+\mu_{ijk}T_{jk}+L_{in}\lambda_n,
%\frac{dy_i}{dt}&=&D_{ij}x_j+E_{ij}y_j+F_{ijk}T_{jk}+L^{(2)}_{in}\lambda_n,\\\label{gen inter3}
%\frac{dT_{ij}}{dt}&=&G_{ijk}x_k+J_{ijk}y_k+K_{ijkl}T_{kl}+L^{(12)}_{ijn}\lambda_n,\\\label{gen inter3}
%\frac{d\lambda_n}{dt}&=&S_{nm}\lambda_m+Z_{ni}x_{i}+P_{ni}y_{i}+V_{nij}T_{ij},\\\label{gen inter2}
\end{eqnarray}
where $a_{ij},b_{ij},\mu_{ijk},L_{in}$ are the components of the generator $H$. Similarly, one can write the differential equation for $\frac{dy_i}{dt},\frac{dT_{ij}}{dt}$ and $\frac{d\lambda_n}{dt}$.

The small spin in the state $\bx(t)$ is assumed to interact via pairwise interaction with each of $N$ spins constituting the large spin. The dynamical equation for the small spin is given by:
\begin{equation}\label{pairwise dynamics}
\frac{dx_i}{dt}=a_{ij}x_j+\sum_{s=1}^N\left( b^{(s)}_{ij}y^{(s)}_j+\mu^{(s)}_{ijk}T^{(s)}_{jk}+L^{(s)}_{in}\lambda^{(s)}_n\right),
\end{equation}
where $\by^{(s)}$ is the Bloch vector of the $s$-th spin of the large system, $T^{(s)}$ is the correlation tensor of the small spin and the $s$-th spin and   $\Lambda^{(s)}=(\dots,\lambda^{(s)}_n,\dots)$ is the set of global parameters of the small spin and all spins of the large one.

We assume that each of the $N$ constituents of the large system interacts with the small spin in a ``same way'', the only difference being in the strength of interaction (for example, because one spin is physically closer to the small spin than the other one.). Thus, one has
\begin{equation}
b^{(s)}_{ij}=\beta_sb_{ij}~~~~\mu^{(s)}_{ijk}=J_s\mu_{ijk},
\end{equation}
where $\beta_s$ and $J_s$ are the coupling constants defining the strength of interaction (they can be different due to the spatial distribution of particles that constitute the large system). Here $b_{ij}$ and $\mu_{ijk}$ are constants that are characteristic of the pairwise interaction and they are assumed to be the same for all particles.

If we assume that in the macroscopic limit, the state of the large system changes negligibly during the interaction time, we obtain
\begin{equation}
T^{(s)}_{ij}(t)=x_i(t)y_j^{(s)}(0),~ \lambda^{(s)}_n(t)=\lambda^{(s)}_n(0)=0, ~ y^{(s)}_j(t)=y^{(s)}_j(0).
\end{equation}
The equation \eqref{pairwise dynamics} becomes:
\begin{eqnarray}\nonumber
\frac{dx_i(t)}{dt}&=&\left( a_{ij}+\mu_{ijk}\sum_{s=1}^N J_sy_k^{(s)}(0)\right)x_j(t)+b_{ij}\sum_{s=1}^N \beta_sy^{(s)}_j(0).
\end{eqnarray}
If $b_{ij}=0$ (otherwise, the equation above does not represent unitary dynamics), this equation becomes equivalent to Eq. \eqref{diff transform} in the limit of very large $N$, in which case one obtains:
\begin{equation}\label{generator}
g_{ij}=a_{ij}+\mu_{ijk}B_k.
\end{equation}
Here $\mathbf{B}=\sum_{s=1}^N J_s\by^{(s)}(0)$ can be understood as an analog of macroscopic field or magnetization, resembling the field produced by a ferromagntic in quantum mechanics. Assuming that the large system is in a spin-coherent state $\vec{n}$, one obtains the same expression as in the case of quantum mechanics: $\vec{B}=N\langle J\rangle\vec{n}$, where $\langle J\rangle=\frac{1}{N}\sum_{i=1}^NJ_n$.

\section{Covariant interaction}

The dynamical equation that follows from \eqref{generator} reads
\begin{equation}\label{dynamics law}
\frac{dx_i}{dt}=(a_{ij}+\mu_{ijk}B_k)x_j,
\end{equation}
where $B_k$ is the component of the macroscopic field. Since we want the dynamics to be reversible (and therefore to transform pure states into pure states), the equation above should preserve the norm of $\bx$. Therefore one has:
\begin{eqnarray}\label{consraint}
a_{ij}=-a_{ji}~~\mbox{and}~~\mu_{ijk}=-\mu_{jik}.
\end{eqnarray}

The dynamics is solely generated by the field $B_k$, since  $a_{ij}$ and $\mu_{ijk}$ are constants that arise from the pairwise dynamics \eqref{gen inter1}.

Let $\mathcal{G}$ be the group of transformations of a single spin (see Section \ref{Single Gbit}). Since there is no external reference direction we assume that the dynamical law \eqref{dynamics law} is covariant, i.e. it has the same form in all frames of reference. More precisely, for any reversible transformation $R\in\mathcal{G}$ (note that $R$ is a transformation on a sphere $\mathcal{S}^{(d-1)}$, therefore it is real and orthogonal $RR^{T}=\openone$) that maps old coordinates of the spin and field into the new ones, $x'_i=R_{ii_{1}}x_{i_1}$ and $B'_i=R_{ii_{1}}B_{i_1}$, we assume that the dynamical law keeps the same form in new coordinates:
\begin{equation}\label{new dynamics law}
\frac{dx'_i}{dt}=(a_{ij}+\mu_{ijk}B'_k)x'_j.
\end{equation}
The tensors $a_{ij}$ and $\mu_{ijk}$ do not change because they are constants of interaction. After the substitution one obtains:
\begin{equation}
R_{ii_1}\frac{dx_{i_1}}{dt}=(a_{ij}+\mu_{ijk}R_{kk_1}B_{k_1})R_{jj_1}x_{j_1}.
\end{equation}
If we multiply the last equation with $R^{-1}=R^{T}$ we obtain
\begin{equation}
\frac{dx_{i_1}}{dt}=(a_{ij}R_{ii_1}R_{jj_1})x_{j_1}+(\mu_{ijk}R_{ii_1}R_{jj_1}R_{kk_1})B_{k_1}x_{j_1}.
\end{equation}

Therefore, for all $R\in\mathcal{G}$ we require:
\begin{eqnarray}\label{covA}
R_{ii_1}R_{jj_1}a_{i_1j_1}&=&a_{ij},\\\label{covE}
R_{ii_1}R_{jj_1}R_{kk_1}\mu_{i_1j_1k_1}&=&\mu_{ijk}.
\end{eqnarray}
Under these conditions, the pairwise interaction \eqref{gen inter1} is invariant under simultaneous change of the local reference frames.

If $\mu_{ijk}=0$, then the dynamics given by \eqref{dynamics law} becomes trivial as it does not depend on the internal state of the transformation device but only on the interaction constant $a_{ij}$ (i.e. the set of transformations becomes one-parameter Lie group). We require that equation \eqref{covE} has non-trivial solution $\mu_{ijk}\neq0$.

\section{Main proofs}
Here we show that only $d=3$ gives non-trivial solution of the equations \eqref{covA} and \eqref{covE}. Recall that the group of physical transformations $\mathcal{G}$ contains the minimal group transitive on the sphere $\mathcal{S}^{(d-1)}$. All such groups are summarized in the Appendix \ref{App_GT on Spheras}.

\subsection{Hint to representation theory}

Our result is based on the group representation theory. We therefore first introduce some basic notions of the representation theory. For an abstract group $\mathcal{G}$ and element $g\in\mathcal{G}$ we say that a matrix $D(g)\in\mathrm{Mat}(\mathcal{H})$, where $\mathcal{H}$ is a vector space, defines a representation of $\mathcal{G}$ if $D(g_1g_2)=D(g_1)D(g_2)$ for every two group elements $g_1$ and $g_2$. In this work we consider only unitary (orthogonal) representations. Representation is called reducible if there exists a nontrivial invariant subspace for all the matrices $D(g)$. Otherwise it is irreducible (IR) representation. Therefore, the group induces a decomposition of the vector space $\mathcal{H}=\oplus_{\mu}\mathcal{H}^{(\mu)}$ into irreducible subspaces $\mathcal{H}^{(\mu)}$ and
\begin{equation}
D(g)=\oplus_{\mu}a_{\mu}\Delta^{(\mu)}(g),
\end{equation}
where $\Delta^{(\mu)}(g)$ is an IR representation that appears with the frequency $a_{\mu}$. The dimension of the IR subspace is $|\mathcal{H}^{(\mu)}|=|\mu|a_{\mu}$, where $|\mu|$ is the dimension of the IR representation $\Delta^{(\mu)}$. The frequency of some IR representation can be computed as
\begin{equation}\label{frequency}
a_{\mu}=(\chi^{(\mu)},\chi)=\frac{1}{|\mathcal{G}|}\sum_{g\in\mathcal{G}}\chi^{(\mu)}(g^{-1})\chi(g),
\end{equation}
where $\chi(g)=\mathrm{Tr}(D(g))$ and $\chi^{(\mu)}(g)=\mathrm{Tr}(\Delta^{(\mu)}(g))$ are the characters of the representations.

For two representations $D_1(\mathcal{G})$ and $D_2(\mathcal{G})$ one can define the tensor product $(D_1\otimes D_2)(\mathcal{G})$ that is representation of $\mathcal{G}$ itself. If $D_{1}$ and $D_2$ are IR, then the decomposition of $D_1\otimes D_2$ is called Clebsch–Gordan (CG) series. In this work, it will be of  particular interest to compute the frequency of the trivial representation $\Delta^{(1)}(g)=1$. The following lemma will be used (see Appendix for the proof):

\begin{Lemma}\label{decomposition lemma} CG series of the product $\Delta^{(\mu)}\otimes\Delta^{(\nu)}$, where $\Delta^{(\mu)},\Delta^{(\nu)}$ are real and irreducible, contains the trivial representation if and only if $\mu=\nu$ and then the trivial representation appears once, only.
\end{Lemma}

The main purpose of introducing the tools of representation theory is to solve Eqs.~\eqref{covA} and \eqref{covE}. The left hand side of Eqs.~\eqref{covA} and \eqref{covE} can be seen as an action of the Kronecker products $D(\mathcal{G})\otimes D(\mathcal{G})$ and $D(\mathcal{G})\otimes D(\mathcal{G})\otimes D(\mathcal{G})$, respectively, with $D(\mathcal{G})$ being the representation of the group of transformation $\mathcal{G}$ and $D(R)=R\in\mathcal{G}$. The solutions $a_{ij}$ and $\mu_{ijk}$ are invariant under the action of the group of transformations $\mathcal{G}$, hence they lie within the totally invariant IR subspace that belongs to the trivial representation. Therefore, we will need the CG decomposition of $D\otimes D$ and $D\otimes D\otimes D$ in order to solve equations \eqref{covA} and \eqref{covE}.

\subsection{$d$ odd case and $d\neq7$}

If we assume $d$ odd, $d>1$ and $d\neq7$, the set of physical transformations contains the special orthogonal group $\mathrm{SO}(d)\triangleleft\mathcal{G}$ (see Appendix~\ref{App_GT on Spheras}).

The easiest way to solve Eq.~\eqref{covA} is as follows. We rewrite it into a matrix form: $RAR^{\TT}=A$ for all $R\in\mathrm{SO}(d)$. This is possible only if $A$ is a scalar matrix $A=a\openone$. Taking into account Eq.~\eqref{consraint} we conclude $A=0$. Although this gives the solution in the particular case considered, we will proceed with full group-representation analysis of Eqs. \eqref{covA} and \eqref{covE} as we will need it later on.

The set of $d\times d$ orthogonal matrices of the unit determinant define a $d$-dimensional, real IR representation of $\mathrm{SO}(d)$ and we label it as $\Delta^{d}$ with $\Delta^{d}(R)=R$. The left-hand side of Eqs.~\eqref{covA} and \eqref{covE} can be seen as the action of product representation $\Delta^{d}(R)\otimes\Delta^{d}(R)$ and $\Delta^{d}(R)\otimes\Delta^{d}(R)\otimes\Delta^{d}(R)$. The solutions $a_{ij}$  and $\mu_{ijk}$ lie within the IR subspace that belongs to the trivial representation in CG series of $\Delta^{d}\otimes\Delta^{d}$ and $\Delta^{d}\otimes\Delta^{d}\otimes\Delta^{d}$, respectively.

Let us analyze the product $\Delta^{d}\otimes\Delta^{d}$. Note that this representation commutes with the permutation group $S_2$ (of two elements). Therefore $\Delta^{d}\otimes\Delta^{d}$ can be decomposed on invariant subspaces that are irreducible under the action of $S_2$. There are two of them and they define symmetric and antisymmetric subspace of the dimensions $\frac{1}{2}d(d+1)$ and $\frac{1}{2}d(d-1)$ spanned by Hermitian and skew-Hermitian matrices, respectively. Furthermore, the symmetric subspace can be decomposed into one-dimensional subspace $\mathcal{H}^{(1)}$ spanned by the identity matrix $\openone$ (invariant under $\Delta^d\otimes\Delta^d$, hence belongs to the trivial subspace) and its orthogonal complement $\mathcal{H}^{(S)}$ of dimension $\frac{1}{2}d(d+1)-1=\frac{1}{2}(d-1)(d+2)$. This induces the decomposition $\mathbb{R}^d\otimes\mathbb{R}^d=\mathcal{H}^{(1)}\oplus\mathcal{H}^{(S)}\oplus\mathcal{H}^{(AS)}$. Each subspace is irreducible for $\Delta^{d}\otimes\Delta^{d}$ (see Appendix \ref{App_SOdxSOd Dec}). Therefore, the CG series is given by:
\begin{equation}\label{SOd decomposition}
\Delta^{d}\otimes\Delta^{d}=\Delta^{1}\oplus\Delta^{S}\oplus\Delta^{AS},
\end{equation}
where $\Delta^{S}$ and $\Delta^{AS}$ are the corresponding IR representations of $\mathrm{SO}(d)$ of the dimensions $AS=\frac{1}{2}d(d-1)$ and $S=\frac{1}{2}(d-1)(d+2)$, respectively. Since the trivial representation appears once only, the solution to \eqref{covA} is one-dimensional and spanned by identity matrix $a_{ij}=a \delta_{ij}$. By applying condition \eqref{consraint} we get $a_{ij}=0$.

The degeneracy of the solution of Eq.~\eqref{covE} can be found from the decomposition
\begin{eqnarray}
\Delta^{d}\otimes\Delta^{d}\otimes\Delta^{d}&=&\Delta^{d}\otimes(\Delta^{d}\otimes\Delta^{d})\\\nonumber
&=&\Delta^{d}\otimes\left(\Delta^{1}\oplus\Delta^{S}\oplus\Delta^{AS}\right)\\\nonumber
&=&\Delta^{d}\otimes\Delta^{1}\oplus\Delta^{d}\otimes\Delta^{S}\oplus\Delta^{d}\otimes\Delta^{AS}.
\end{eqnarray}

Let us apply lemma \ref{decomposition lemma}. The decomposition of the first term $\Delta^{d}\otimes\Delta^{1}$ in last equation does not contain the trivial representation because $d>1$. The second term $\Delta^{d}\otimes\Delta^{S}$ contains the trivial representation, if $\Delta^{d}$ and $\Delta^{S}$ are equivalent, which is possible only if $d=S=\frac{1}{2}(d-1)(d+2)$. There is no solution to this equation among $d$ odd numbers. Similarly, the last term contains the trivial representation, only if $d=\frac{1}{2}d(d-1)$, that has solution $d=3$. Furthermore for $d=3$ the solution is one-dimensional and is represented by the completely antisymmetric (Levi-Civita) tensor $\mu_{ijk}=\epsilon_{ijk}$, where $\epsilon_{ijk}=+1$ for $(ijk)$ being an even permutation. The solution $d=3$ is worked out in details in Appendix~\ref{App_d=3 Sol}.

\subsection{$d=7$ case}
Here the minimal transitive group on $S^6$ is the exceptional Lie group $G_2$. The generators span a 14-dimensional Lie algebra:
\begin{equation}\label{G2 generator}
H(\bx)=\left(
\begin{smallmatrix}%{ccccccc}
 0 & x_1 & -x_2 & x_3 & -x_4 & -x_5 & x_9-x_7 \\
 -x_1 & 0 & x_6 & x_7 & x_8-x_5 & x_4-x_{11} & x_3+x_{10} \\
 x_2 & -x_6 & 0 & -x_8 & x_9 & x_{10} & x_{11} \\
 -x_3 & -x_7 & x_8 & 0 & x_{13}-x_6 & x_{14}-x_2 & x_{12}-x_1 \\
 x_4 & x_5-x_8 & -x_9 & x_6-x_{13} & 0 & x_{12} & -x_{14} \\
 x_5 & x_{11}-x_4 & -x_{10} & x_2-x_{14} & -x_{12} & 0 & x_{13} \\
 x_7-x_9 & -x_3-x_{10} & -x_{11} & x_1-x_{12} & x_{14} & -x_{13} & 0
\end{smallmatrix}
\right).
\end{equation}

We will next show that this generator, in general is not of the form \eqref{generator}, i.e. $H_{ij}=a_{ij}+\mu_{ijk}B_k$. This means that the dynamics generated by macroscopic field $B_k$ exceeds the group $G_2$. On the other hand, there is no group transitive on $S^6$ other than $G_2$ and $\mathrm{SO}(7)$ (see Appendix~\ref{App_GT on Spheras}). Since the group of transformations exceeds $G_2$, it has to be $\mathrm{SO}(7)$. But the case of $\mathrm{SO}(7)$ has been studied in the previous section, where it was shown that no nontrivial solution to Eq.~\eqref{covE} exists in this case.

We apply the analysis from the last section in the present case. We label $7$-dimensional IR representation of $G_2$ as $\Delta^7$. According to Behrends \emph{et al.}~\cite{Behrends} CG series is given by

\begin{equation}\label{G2 decomposition}
\Delta^{7}\otimes\Delta^{7}=\Delta^{1}\oplus\Delta^{7}\oplus\Delta^{14}\oplus\Delta^{27},
\end{equation}
hence the trivial representation appears once only. Consequently the solution to \eqref{covA} is spanned by the identity matrix $a_{ij}=a\delta_{ij}$. Constraint \eqref{consraint} gives $a_{ij}=0$. Next, in the decomposition
\begin{eqnarray}
\Delta^{7}\otimes\Delta^{7}\otimes\Delta^{7}&=&\Delta^{7}\otimes(\Delta^{7}\otimes\Delta^{7})\\\nonumber
&=&\Delta^{7}\otimes\left(\Delta^{1}\oplus\Delta^{7}\oplus\Delta^{14}\oplus\Delta^{27}\right)\\\nonumber
&=&\Delta^{7}\otimes\Delta^{1}\oplus\Delta^{7}\otimes\Delta^{7}\oplus\Delta^{7}\otimes\Delta^{14}\oplus\Delta^{7}\otimes\Delta^{27},
\end{eqnarray}
the trivial representation appears once only due to the term $\Delta^{7}\otimes\Delta^{7}$. Therefore, the solution of equation \eqref{covE} is unique (up to a constant) and is given by completely antisymmetric tensor $\psi_{ijk}$ taking the non-zero value of $+1$ for $ijk$=123, 145, 176, 246, 257, 347, 365. Incidentally, note that $\psi_{ijk}$ is the tensor involved in the definition of the multiplication rule of octonions and seven-dimensional cross product~\cite{Baez}:
\begin{equation}
(\ba\times\bb)_i=\psi_{ijk}a_jb_k,
\end{equation}
where $\ba$ and $\bb$ are two octonions.

Let us set the macroscopic field of \eqref{generator} to $B^{(1)}_k=B\delta_{1k}$. The corresponding generator $g_{ij}=\psi_{ijk}B^{(1)}_k=B\psi_{ij1}$ has six nonzero elements $g_{23}=g_{45}=g_{76}=-g_{32}=-g_{54}=-g_{67}=+1$:

\begin{equation}
G=B\left(
\begin{smallmatrix}%{ccccccc}
 0 & 0 & 0 & 0 & 0 & 0 & 0 \\
 0 & 0 & 1 & 0 & 0 & 0 & 0 \\
 0 & -1 & 0 & 0 & 0 & 0 & 0 \\
 0 & 0 & 0 & 0 & 1 & 0 & 0 \\
 0 & 0 & 0 & -1 & 0 & 0 & 0 \\
 0 & 0 & 0 & 0 & 0 & 0 & -1 \\
 0 & 0 & 0 & 0 & 0 & 1 & 0
\end{smallmatrix}
\right).
\end{equation}

This generator is not of the form~\eqref{G2 generator}; therefore the dynamics generated by $B^{(1)}_k$ goes beyond the $G_2$ group. Since the only transitive groups on $S^6$ are $G_2$ and $\mathrm{SO}(7)$, and since we already excluded $\mathrm{SO}(7)$ in previous section, the Eq.~\eqref{covE} has no solution.

\subsection{$d=4k$ case ($k=\frac{1}{2},1,2,\dots$)}

In this case the minimal group transitive on $\mathcal{S}^{d-1}$ contains total inversion $E\bx=-\bx$. Now, we set $R=E$ in the equation \eqref{covE}, hence $-\mu_{ijk}=\mu_{ijk}$. This gives only trivial solution $\mu_{ijk}=0$.

\subsection{$d=4k+2$ case ($k=1,2,3,\dots$)}

In this case the minimal transitive group is $\mathrm{SU}(2k+1)$. For some complex unitary $u\in\mathrm{SU}(2k+1)$ its representation in $(d=4k+2)$-dimensional real space is given by the following matrix:
\begin{equation}
D(u)=\left(
       \begin{array}{cc}
         \mathrm{Re}~u & -\mathrm{Im}~u \\
         \mathrm{Im}~u & \mathrm{Re}~u \\
       \end{array}
     \right).
\end{equation}
Note that this representation commutes with the symplectic form $J=\left(
                     \begin{smallmatrix}
                       0 & \openone \\
                       -\openone & 0 \\
                     \end{smallmatrix}
                   \right)$, i.e.
\begin{equation}
[D(u),J]=0,
\end{equation}
for every $u$.

Let us analyze the case where $u$ is a real matrix, that is $u\in\mathrm{SO}(2k+1)\triangleleft\mathrm{SU}(2k+1)$. Then $D(u)=\openone_2\otimes u$, where $\openone_2$ is a $2\times2$ identity matrix. The equation \eqref{covE} can be written in a tensor form:
\begin{equation}
(\openone_2\otimes u)\otimes(\openone_2\otimes u)\otimes(\openone_2\otimes u)\ket{\mu}=\ket{\mu},
\end{equation}
or equivalently
\begin{equation}\label{sold6}
(\openone_2\otimes\openone_2\otimes\openone_2)\otimes (u\otimes u\otimes u)\ket{\widetilde{\mu}}=\ket{\widetilde{\mu}},
\end{equation}
where $\ket{\widetilde{\mu}}$ and $\ket{\mu}$ are the ket vectors that correspond to the tensors $\mu_{ijk}$ and $\widetilde{\mu}_{ijk}$ and are connected by a suitable transformation. The solution to the last equation can be found in a product form $\ket{\widetilde{\mu}}=\ket{\chi}\ket{\phi}$, where
\begin{equation}
(u\otimes u\otimes u)\ket{\phi}=\ket{\phi},
\end{equation}
holds for every $u\in\mathrm{SO}(2k+1)$. This equation has been analyzed earlier and it has nontrivial solution only if $2k+1=3$ or $d=6$. In that case, solution $\ket{\phi}$ has components $\phi_{ijk}$  that are the Levi-Civita tensor $\epsilon_{ijk}$, hence we write the solution as $\ket{\widetilde{\mu}}=\ket{\chi}\ket{\epsilon}$ .

We have found non-trivial solution for the case $d=6$ and the corresponding group is $\mathrm{SU}(3)$. The group generators span 8-dimensional Lie algebra and the corresponding real representation reads:
\begin{equation}\label{SU3 generator}
H(\bx)=\left(
\begin{array}{cccccc}
 0 & -x_4 & -x_5 & x_7 & x_1 & x_2 \\
 x_4 & 0 & -x_6 & x_1 & x_8-x_7 & x_3 \\
 x_5 & x_6 & 0 & x_2 & x_3 & -x_8 \\
 -x_7 & -x_1 & -x_2 & 0 & -x_4 & -x_5 \\
 -x_1 & x_7-x_8 & -x_3 & x_4 & 0 & -x_6 \\
 -x_2 & -x_3 & x_8 & x_5 & x_6 & 0
\end{array}
\right).
\end{equation}
We set the notation $H_i=H(\be^{(i)})$, where $e^{(i)}_k=\delta_{ik}$ is the $k$th component of  $e^{(i)}_k$.

Similarly to the previous section our goal is to show that $a_{ij}+\mu_{ijk}B_k$ generate transformations that go beyond the $\mathrm{SU}(3)$ group. In such a case, the group of transformations exceeds the minimal transitive group. Since there is no group transitive on $S^5$ other than $\mathrm{SU}(3)$ that do not contain the total inversion, one concludes that there is no nontrivial solution to Eq.~\eqref{covE}.

Note that the solution to Eq.~\eqref{covA} is twofold $a_{ij}=\alpha\delta_{ij}+\beta J_{ij}$, where $J_{ij}$ is the symplectic form. However, since $a_{ij}=-a_{ji}$ we have $\alpha=0$. Furthermore, symplectic form $J_{ij}$ does not belong to the set of generators $H(\bx)$ therefore $\beta=0$ and finally $a_{ij}=0$.

Recall that the solution to \eqref{sold6} can be found in the product form $\ket{\chi}\ket{\epsilon}$ where $\ket{\epsilon}$ is the tensor Levi-Civita. Let us set the macroscopic field of \eqref{generator} to $B^{(1)}_k=B\delta_{1k}$. In that case the generator becomes $G=B\chi\otimes E_1$, where $\chi_{ab}$ is some symmetric $2\times2$ matrix and  $[E_1]_{ij}=\epsilon_{ij1}$. One has
\begin{equation}
G=B\left(
    \begin{array}{cccccc}
      0 & 0 & 0 & 0 & 0 & 0 \\
      0 & 0 & \chi_{11} & 0 & 0 & \chi_{12} \\
      0 & -\chi_{11} & 0 & 0 & -\chi_{12} & 0 \\
      0 & 0 & 0 & 0 & 0 & 0 \\
      0 & 0 & \chi_{21} & 0 & 0 & \chi_{22} \\
      0 & -\chi_{21} & 0 & 0 & -\chi_{22} & 0 \\
    \end{array}
  \right).
\end{equation}
This can be generator of the form \eqref{SU3 generator} if $\chi_{12}=\chi_{21}=0$ and $\chi_{11}=\chi_{22}=\chi_0$. Therefore $G=B\chi_0 H_6$. On the other hand, the dynamics generated by $B^{(1)}_k=\delta_{k1}$ is invariant under all transformations that keep $B^{(1)}_k$ invariant. In this particular case, it means that $G$ has to commute with the generators $H_6$ and $H_3$. This gives only trivial solution $\chi_0=0$, hence $G=0$. Similarly, one can draw the same conclusion for any other $B^{(s)}_k=B\delta_{ks}$. Therefore, the dynamics generated by arbitrary $B_k$ goes beyond the $\mathrm{SU}(3)$ group.

\section{Going beyond three dimensions}\label{Going beyond d=3}

In this section we shall argue that higher-dimensional macroscopic limit may arise as a consequence of a multi-partite invariant interaction among elementary spins. We construct an explicit model of dynamics in analogy to the quantum case and three dimensions (see Appendix \ref{App_d=3 Sol} for details). However, it remains as an open question if such an ansatz leads to a proper probabilistic theory, in the sense that positivity of probabilities is not guaranteed.

Note that it as an artefact of the three dimensions that the evolution equation~\eqref{diff transform} can be written in the form
\begin{equation}
\frac{d\bx}{dt}=\bB\times\bx,
\end{equation}
where the vector $\bB$ generates evolution with the generator matrix $g_{ij}=\epsilon_{ijk}B_k$. This expression for $d>3$ is no longer possible. The evolution cannot be generated by a single vector, but a tensor. We will show that such a situation arises in the macroscopic limit if elementary interactions were multi-particle.

Let us start with the dimension $d=4$. We consider three generalized spins described by a state
\begin{equation}
\psi=\{\bx,\by,\bz,T^{(12)},T^{(13)},T^{(23)},T^{(123)},\Lambda\}.
\end{equation}
Let the spins interact via genuine three-particle, rotationally invariant interaction (see Figure~\ref{Spin Coherent Rotation}, rigth). In analogy with the quantum case discussed above, we can consider the dynamical equation for, say the first spin, as follows:
\begin{equation}\label{three inter}
\frac{dx_i}{dt}=a\epsilon_{ijkl}T_{jkl}^{(123)}+L^{(1)}_{in}\lambda_n.
\end{equation}
Here, $a$ is a constant and $\epsilon_{ijkl}$ is the completely antisymmetric tensor of four indices, with $\epsilon_{1234}=+1$. It is well know that this tensor is invariant under $\mathrm{SO}(4)$ rotations. Analogously, one can write the equations for the other two local Bloch vectors $y_i$ and $z_i$, as well as for correlations, both bipartite and tripartite and the global parameter.

Next we consider an ensemble of a large number $N$ of spins. Let a single spin interact with each of the $N$ spins via three-partite interaction defined above. In the macroscopic limit the dynamics should factorize and the state of the large system of $N$ spins should not evolve in time. Therefore, all the correlations between single spin and large system factorize:
\begin{eqnarray}
T^{(0nm)}_{ijk}(t)&=&x_{i}(t)T^{(nm)}_{jk}(0),\\\nonumber
T^{(0m)}_{ij}(t)&=&x_{i}(t)y^{(m)}_{j}(0),\\\nonumber
\Lambda(t)&=&0,
\end{eqnarray}
where index $0$ labels the single spin, whereas $n$ labels the $n$th spins of the large system ($n=1\dots N$). The $\Lambda$ labels the set of all global parameters between the single spin and the large system.

The equation of motion for the single spin reads:
\begin{equation}\label{three dynamics}
\frac{dx_i}{dt}=a\epsilon_{ijkl}x_j\sum_{n,m=1}^NJ_{nm}T^{(nm)}_{kl}(0),
\end{equation}
where $J_{nm}$ is the coupling constant between the single spin and spins $n$ and $m$ of the large system. Taking $B_{ij}=a\sum_{n,m=1}^NJ_{nm}T^{(nm)}_{kl}(0)$ one obtains a reversible dynamics of a single spin:
\begin{equation}
\frac{dx_i}{dt}=\epsilon_{ijkl}B_{kl}x_j,
\end{equation}
The dynamics is then generated by a covariant tensor field $B_{ij}$. We can further assume the situation as described in Figure~\ref{Spin Coherent Rotation}, right. The spins of the large system are arranged in a (regular) lattice such that each cell consist of two spins prepared along orthogonal directions $\vec{n}_1$ and $\vec{n}_2$. The two arrays of spins define two spin-coherent states. If we assume that the small spin interacts with two spins of a single cell, we obtain $B_{ij}=N\langle J\rangle n_{1i}n_{2j}$, where $\langle J\rangle=\frac{1}{N}\sum_{n=1}^{N}J_n$. We can say that dynamics is generated by two spin-coherent states defined by directions $\vec{n}_1$ and $\vec{n}_2$.

The present analysis for $d=4$ can be generalized to higher-dimensions in a straightforward way. The dynamics of a generalized spin in $d$ dimensions can be obtained from the $\mathrm{SO}(d)$ invariant dynamics that is generated by a genuine $(d-1)$-particle interaction. Of course, it is an open question if the set of equations \eqref{three inter} leads to a proper physical solution, in the sense that positivity of probabilities is not violated. We leave this question open for future investigation.

\section{Conclusions}

Physicist study models with extra dimensions. This research appears to be justified as we do not know of convincing arguments why we should necessarily live in three-dimensional space (or 3+1 space-time). In this paper we put a ``closeness'' requirement on every physical theory, which restricts the possible dimensions.
The theory is closed if macroscopic field - which, via interaction with a microscopic system generates its dynamics - itself is described by the theory in the classical limit.

In the operational approach to a physical theory, one expects that the dimension and the symmetry of the state space of the ``elementary system'' are the same as those of the space in which ``laboratory devices'' are embedded. This is for the simple reason that the parameters describing the state operationally have no other meaning than that of the parameters that specify the configuration of macroscopic instruments by which the states are prepared, transformed or measured. On the other hand, the states of the macroscopic instruments can be obtained from within the theory in the classical limit; for example, in quantum mechanics, the ``magnetic field'' is represented by the coherent state of a very large number of equally prepared spins. Arbitrary unitary transformation of the elementary quantum spin (spin-$1/2$) can be generated by a (group invariant) bipartite interaction between the spin and the ``magnetic field'' (i.e. between the spin and each of the spins constituting the coherent state that represents the ``field''). Therefore quantum theory is closed according to our requirement.

We showed that in no probabilistic theory of spin (where the spin has $d$ components), other than quantum mechanics ($d=3$), an invariant \emph{pairwise} interaction can generate the group of transformation of the spin. However, if one considers three- or more-spin interactions this possibility might be realized. This opens up a possibility of having higher-dimensional spaces ($d>3$) and ``laboratory devices'' embedded in it, which could generate the group of transformation of spin with the state space dimension $d>3$.  We hope that our work will be useful for physicists considering the existence of extra dimensions or other modifications of space-time.

\appendix

\section{Dynamics of spin in presence of spin-coherent state}\label{App_Spin_Dyn}

Here we justify the approximation made in Section \ref{Section: dynamics and macro limit}. Namely, we show that equation~\eqref{generator} can be realized within quantum mechanics. We follow the idea given in the work by Poulin~\cite{Poulin1}. Let the large system be a ferromagnet composed of $N$ spin-$1/2$ particles with the Hamiltonian $H_0$. We assume that $H_0$ is rotationally invariant $U^{\otimes N}H_0 U^{\dagger\otimes N}=H_0$ for all single particle rotations $U\in\mathrm{SU}(2)$. One particular example of such a system is a Heisenberg ferromagnet with the Hamiltonian:
\begin{equation}
H_0=-\sum_{n,m=1}^NJ_{nm}\vec{\sigma^{(n)}}\vec{\sigma^{(m)}},
\end{equation}
with  $J_{nm}\geq0$ are the coupling constants. The rotational invariance is an important assumption because there is no external reference direction. The large system itself can be used to define preferred direction in space. Referring to the well known result in solid state physics~\cite{AshroftMermin} such a system, although rotationally invariant, can still exhibit spontaneous magnetization bellow the critical temperature. At zero temperature all the spins are aligned along some direction, that we choose to be the $\be_z$-direction. Hence the ground state is $\ket{\psi_0}=\ket{0}^{\otimes N}$ with the energy set to zero $E_0=0$ (this is always possible by changing the energy reference point). Let the small system be prepared in a state $\ket{\phi}=\alpha\ket{0}+\beta\ket{1}$ and assume $\sigma_3\ket{0}=\ket{0}$. The system interacts with the large system via Heisenberg interaction, therefore the total Hamiltonian reads
\begin{equation}
H=\sum_{n=1}^NJ_n\vec{\sigma^{(0)}}\vec{\sigma^{(n)}}+H_0,
\end{equation}
where $J_n$ is the coupling constant for the interaction between the small spin and $n$th spin of the large system. Our goal is to show that in macroscopic limit $N\rightarrow\infty$, the dynamics becomes separable:
\begin{eqnarray}\label{eff dynamics}
e^{itH}\ket{\phi}\ket{\psi_0}&=&(e^{itH_{eff}}\ket{\phi})\ket{\psi_0},
\end{eqnarray}
where $H_{eff}$ is an effective Hamiltonian.

Firstly, let us compute the following
\begin{eqnarray}\label{Hpsipsi0}
H\ket{\phi}\ket{\psi_0}&=&\sum_{n=1}^N(J_n\vec{\sigma^{(0)}}\vec{\sigma^{(n)}}+H_0)\ket{\phi}\ket{0}^{\otimes N}\\\nonumber
&=&\sum_{n=1}^NJ_n\vec{\sigma^{(0)}}\vec{\sigma^{(n)}}\ket{\phi}\ket{0}^{\otimes N}\\\nonumber
&=&(\sum_{n=1}^NJ_n)(\sigma_3\ket{\phi})(\sigma_3^{(n)}\ket{0}^{\otimes N})\\\nonumber
&+&\sum_{n=1}^{N}J_n\sum_{i=1}^2\sigma_i^{(0)}\sigma_i^{(n)}\ket{\psi}\ket{0}^{\otimes N}\\\nonumber
&=&(\sum_{n=1}^NJ_n)(\sigma_3\ket{\phi})\ket{0}^{\otimes N}+\sum_{n=1}^{N}J_n\sum_{i=1}^2\sigma_i^{(0)}\sigma_i^{(n)}\ket{\psi}\ket{0}^{\otimes N}\\\nonumber
&=&\ket{\chi}+\ket{\mu},
\end{eqnarray}
where
\begin{eqnarray}
\ket{\chi}&=&(\sum_{n=1}^NJ_n)(\sigma_3\ket{\phi})\ket{0}^{\otimes N},\\
\ket{\mu}&=&\sum_{n=1}^{N}J_n\sum_{i=1}^2\sigma_i^{(0)}\sigma_i^{(n)}\ket{\psi}\ket{0}^{\otimes N}.
\end{eqnarray}
The norm of $\ket{\chi}$ is easy to compute $\braket{\chi}{\chi}=(\sum_{n=1}^NJ_n)^2$. On the other hand, we have:
\begin{eqnarray}
\ket{\mu}&=&(\sigma_1\ket{\psi})(J_1\ket{1}\ket{0}\ket{0}\dots+J_2\ket{0}\ket{1}\ket{0}\dots)\\\nonumber
&+&(i\sigma_2\ket{\psi})(J_1\ket{1}\ket{0}\ket{0}\dots+J_2\ket{0}\ket{1}\ket{0}\dots)\\\nonumber
&=&\left((\sigma_1+i\sigma_2)\ket{\psi}\right)(J_1\ket{1}\ket{0}\ket{0}\dots+J_2\ket{0}\ket{1}\ket{0}\dots).
\end{eqnarray}
The norm of $\ket{\mu}$ is given by $\braket{\mu}{\mu}=\sum_{n=1}^NJ^2_n$. Let us define the averages
\begin{eqnarray}
\langle J\rangle_N&=&\frac{1}{N}\sum_{n=1}^NJ_n,\\
\langle J^2\rangle_N&=&\frac{1}{N}\sum_{n=1}^NJ^2_n.
\end{eqnarray}
We assume that $\langle J\rangle_N$ and $\langle J^2\rangle_N$ have finite values in macroscopic limit. Furthermore, we assume that $\lim_{N\rightarrow\infty}\langle J\rangle_N=\langle J\rangle\neq0$. We can express the norms of $\ket{\chi}$ and $\ket{\mu}$ in terms of these quantities
\begin{eqnarray}
\braket{\chi}{\chi}&=&N^2\langle J\rangle_N,\\
\braket{\mu}{\mu}&=&N\langle J^2\rangle_N.
\end{eqnarray}
Now, it is clear that $\ket{\mu}$ is a vector of short length as compared to $\ket{\chi}$ when $N$ is large. Furthermore, in the macroscopic limit, we have $\lim_{N\rightarrow\infty}\frac{\braket{\mu}{\mu}}{\braket{\chi}{\chi}}=0$, therefore one can safely remove $\ket{\mu}$ from equation \eqref{Hpsipsi0} when $N\rightarrow\infty$:
\begin{equation}
H\ket{\phi}\ket{\psi_0}=(H_{eff}\ket{\phi})\ket{\psi_0},
\end{equation}
where $H_{eff}=N\langle J\rangle\sigma_3$. Now we can prove \eqref{eff dynamics}:
\begin{eqnarray}
e^{itH}\ket{\phi}\ket{\psi_0}&=&\sum_{k=0}^{+\infty}\frac{t^k}{k!}H^k\ket{\phi}\ket{\psi_0}\\\nonumber
&=&\sum_{k=0}^{+\infty}\frac{t^k}{k!}(H_{eff}^k\ket{\phi})\ket{\psi_0}\\\nonumber
&=&(e^{itH_{eff}}\ket{\phi})\ket{\psi_0}.
\end{eqnarray}

In general, if the large system exhibits the ground state $\ket{\psi_0}=\ket{\vec{n}}^{\otimes N}$ (spin coherent state) magnetized along the direction $\vec{n}$, it will generate an effective Hamiltonian $H_{eff}(\vec{n})=N\langle J\rangle\vec{n}\vec{\sigma}$.

\section{Groups Transitive on spheres}\label{App_GT on Spheras}
The groups that are transitive on spheres are summarized in Table~\ref{GT table}.
\begin{table}[h!]
\begin{tabular}{|c|c|}
  \hline
  abstract group & $d$\\\hline
  $\mathrm{SO}(d)$ & $3,4,5,\dots$\\
  $\mathrm{SU}(d/2)$ & $4,6,8,\dots$\\
  $\mathrm{U}(d/2)$ & $2,4,6,8,\dots$\\
  $\mathrm{Sp}(d/4)$ & $8,12,16,\dots$\\
  $\mathrm{Sp}(d/4)\times\mathrm{U}(1)$ & 8,$12,16,\dots$\\
  $\mathrm{Sp}(d/4)\times\mathrm{SU}(2)$ & $4,8,12,\dots$\\
  $\mathrm{G}_2$ & $7$\\
  $\mathrm{Spin}(7)$ & $8$\\
  $\mathrm{Spin}(9)$ & $16$ \\
  \hline
\end{tabular}
\caption{Table taken from the Ref.~[\onlinecite{Massanes}]. We assume $d>1$ always. First column shows the abstract group transitive on sphere $\mathcal{S}^{d-1}$, whereas the second column shows the possible value of $d$. Here  $\mathrm{SO}(2)\cong\mathrm{U}(1)$ and $\mathrm{Sp}(1)\cong\mathrm{SU}(2)$. For a complex matrix $U$, the real representation is generated by following real matrix~\usebox{\smlmat}.}
\label{GT table}
\end{table}

For simplicity reasons, we shall study only the minimal group (therefore certainly within the set of physical transformations) that is transitive on a sphere $\mathcal{S}^{d-1}$. If $d$ is odd, the minimal transitive group is the special orthogonal group $\mathrm{SO}(d)$ unless $d=7$. For $d=7$ the minimal group is the exceptional Lie group $\mathrm{G}_2$. If $d$ is even, there are several options. We distinguish the cases whether the group contains the total inversion $E\bx=-\bx$ or not. The groups $\mathrm{U}(d/2), \mathrm{Sp}(d/4), \mathrm{Sp}(d/4)\times\mathrm{U}(1), \mathrm{Sp}(d/4)\times\mathrm{SU}(2), \mathrm{Spin}(7)$ and $\mathrm{Spin}(9)$ contain $E$ as well as the group $\mathrm{SU}(d/2)$, if $d$ is multiple of four $d=4k$ (Ref.~[\onlinecite{Massanes}], page 18). The only $d$-even groups that do not contain total inversion are $\mathrm{SU}(d/2)$ for $d=4k+2$, where $k=1,2,3,\dots$

\section{Kronecker product of irreducible representations}\label{App_Lemma1}

Here we provide the proof of lemma \ref{decomposition lemma}:

{\bf Lemma 1.~}{\em CG series of the product $\Delta^{(\mu)}\otimes\Delta^{(\nu)}$, where $\Delta^{(\mu)},\Delta^{(\nu)}$ are real and irreducible, contains the trivial representation if and only if $\mu=\nu$ and then the trivial representation appears once, only.}

{\footnotesize {\bf Proof.}
Note that for a real, orthogonal representation $D(g)$ we have $D(g^{-1})=D^{\TT}(g)$, hence $\chi(g^{-1})=\mathrm{Tr}D(g^{-1})=\mathrm{Tr}D^{\TT}(g)=\chi(g)$. We set $\mu=1$ with $\Delta^{(1)}(g)=1$ (trivial representation) and $D(g)=\Delta^{(\mu)}(g)\otimes\Delta^{(\nu)}(g)$. We have the characters $\chi^{(1)}(g)=1$ and $\chi(g)=\chi^{(\mu)}(g)\chi^{(\nu)}(g)$. The frequency is computed using Eq.~\eqref{frequency}
\begin{eqnarray}
a_{1}&=&(\chi^{(1)},\chi)\\
&=&\frac{1}{|\mathcal{G}|}\sum_{g\in\mathcal{G}}\chi^{(1)}(g^{-1})\chi^{(\mu)}(g)\chi^{(\nu)}(g)\\
&=&\frac{1}{|\mathcal{G}|}\sum_{g\in\mathcal{G}}\chi^{(\mu)}(g)\chi^{(\nu)}(g)\\
&=&\frac{1}{|\mathcal{G}|}\sum_{g\in\mathcal{G}}\chi^{(\mu)}(g^{-1})\chi^{(\nu)}(g)\\
&=&(\chi^{(\mu)},\chi^{(\nu)})\\
&=&\delta_{\mu\nu}.
\end{eqnarray}
\begin{flushright}
QED
 \end{flushright}
}

\section{Irreducible decomposition of the two-fold tensor representation of $\mathrm{SO}(d)$}\label{App_SOdxSOd Dec}

Here we show that the decomposition \eqref{SOd decomposition}
\begin{equation}
\Delta^{d}\otimes\Delta^{d}=\Delta^{1}\oplus\Delta^{S}\oplus\Delta^{AS},
\end{equation}
is irreducible unless $d=4$.

Let the representation $D(\mathcal{G})$ of $\mathcal{G}$ acts on a vector space $\mathcal{V}$. By definition, $D(\mathcal{G})$ is irreducible on $\mathcal{V}$ if $\mathrm{span}\{D(g)\bx~|~\forall g\in\mathcal{G}\}=\mathcal{V}$ for every non-zero vector $\bx\in \mathcal{V}$.

Firstly, let us analyze the symmetric subspace of all $d\times d$ symmetric, traceless matrices
\begin{equation}
\mathcal{V}_S=\{H~|~H^{\TT}=H~\wedge~\Tr H=0\}.
\end{equation}
This is an invariant subspace under the action of $\mathrm{SO}(d)$, because $(RHR^{\TT})^{\TT}=RHR^{\TT}$ for every $R\in\mathrm{SO}(d)$ and $H\in\mathcal{V}_S$. Our goal is to show that the action of $\mathrm{SO}(d)$ is irreducible on $\mathcal{V}_S$. Therefore, we have to prove that the set
\begin{equation}
\mathcal{W}(H)=\mathrm{span}\{RHR^{\TT}~|~R\in\mathrm{SO}(d)\}=\mathcal{V}_S,
\end{equation}
for every non-zero $H\in\mathcal{V}_S$. Let us write $H$ in diagonal form $H=\sum_{i=1}^dh_i\ket{i}\bra{i}$, where $H=\sum_{i=1}^dh_i=0$. Since $\Tr H=0$, the largest and lowest eigenvalue satisfy $h_{\max}>0$ and $h_{\min}<0$. For convenience we set $h_{\max}=h_1$ and $h_{\min}=h_2$. Consider the orthogonal matrix $F_{12}\in\mathrm{SO}(d)$ swapping the basis vectors  $\ket{1}$ and $\ket{2}$ (swap-rotation in $12$-subspace):
\begin{equation}
F_{12}=\mathrm{diag}\left[\left(
                  \begin{array}{cc}
                    0 & -1 \\
                    1 & 0 \\
                  \end{array}
                \right),1,1,1,\dots
\right].
\end{equation}
We have $H'=\frac{1}{h_1-h_2}(H-F_{12}HF_{12}^{\TT})=\ket{1}\bra{1}-\ket{2}\bra{2}$, where $h_1-h_2>0$. If we further rotate in $12$-subspace for $45^\circ$ we obtain
\begin{equation}
R_{45^\circ}H'R_{45^\circ}^{\TT}=\ket{1}\bra{2}+\ket{2}\bra{1}=E_{12}.
\end{equation}
The matrix $E_{12}$ is the element of a standard basis in $\mathcal{V}_S$. Other basis elements $E_{ij}$ can be obtained from $E_{12}$ by suitable rotations. Therefore we have completed the space $\mathcal{V}_S$ starting from an arbitrary element $H$, hence $\mathcal{W}_S(H)=\mathcal{V}_S$.

In the case of antisymmetric subspace we define
\begin{equation}
\mathcal{V}_{AS}=\mathrm{span}\{H~|~H^{\TT}=-H\}.
\end{equation}
Our goal is to show $\mathcal{W}(A)=\mathcal{V}_{AS}$ for arbitrary $A\in\mathcal{V}_{AS}$. Let $A_{ij}=\ket{i}\bra{j}-\ket{j}\bra{i}$, for $j>i$ be the standard basis in $\mathcal{V}_{AS}$. It is sufficient to show that $A_{12}\in\mathcal{W}(A)$, and the other basis elements can be obtained from $A_{12}$ by suitable rotations. For an arbitrary antisymmetric matrix $A\in\mathcal{V}_{AS}$ we can find the canonical form by applying suitable rotation $T\in\mathrm{SO}(d)$:
\begin{eqnarray}\nonumber
A'=TAT^{\TT}&=&\mathrm{diag}\left[\left(
                  \begin{array}{cc}
                    0 & -a_1 \\
                    a_1 & 0 \\
                  \end{array}
                \right),\left(
                  \begin{array}{cc}
                    0 & -a_2 \\
                    a_2 & 0 \\
                  \end{array}
                \right),\dots,0,0,\dots\right]\\
&=&a_1A_{12}+a_2A_{34}+\dots.
\end{eqnarray}
If only $a_1\neq0$, than $A=a_1A_{12}$ and we can generate the full basis $\{A_{ij}\}$ in $\mathcal{V}_{AS}$ by applying suitable rotations. Otherwise, we assume that at least two elements $a_i$ are non-zero, and for convenience we set $a_{1}\neq 0$ and $a_2\neq 0$. Let $R_{ij}$ be the rotation that flips $i$th and $j$th coordinate only, i.e. $R_{ij}\ket{k}=s\ket{k}$, where $s=-1$ if $k=i$ or $k=j$, otherwise $s=1$. We get the following
\begin{equation}
A''=A'-R_{13}A'R_{13}^{\TT}=2a_1A_{12}+2a_2A_{34}.
\end{equation}

Now if $d>4$ we further apply $R_{15}$ to $A''$ and obtain the following
\begin{equation}
A''-R_{15}A''R_{15}^{\TT}=4a_1A_{12}.
\end{equation}
From here we can generate the full basis $A_{ij}$, hence $\mathcal{W}(H)=\mathcal{V}_{AS}$. If $d=4$ the construction above is no longer possible ($R_{15}$ does not exist). In this case the antisymmetric space is reduced to two three-dimensional irreducible subspaces as follows
\begin{equation}
\Delta^{4}\otimes\Delta^{4}=\Delta^{1}\oplus\Delta^{9}\oplus\Delta^{3}_{+}\oplus\Delta^{3}_{-}.
\end{equation}
We leave the proof to the curious reader.

\section{$d=3$ solution}\label{App_d=3 Sol}

We begin with analyzing the fourth tensor power of $\Delta^{d}$ representation of $\mathrm{SO}(d)$ group, as defined in the main text. We have
\begin{eqnarray}
\Delta^{d}\otimes\Delta^{d}\otimes\Delta^{d}\otimes\Delta^{d}&=&(\Delta^{d}\otimes\Delta^{d})\otimes(\Delta^{d}\otimes\Delta^{d})\\\nonumber
&=&\left(\Delta^{1}\oplus\Delta^{AS}\oplus\Delta^{S}\right)\otimes\left(\Delta^{1}\oplus\Delta^{AS}\oplus\Delta^{S}\right).
\end{eqnarray}
Since $S\neq AS$ for $d>1$ and $d\neq4$ (see Appendix \ref{App_SOdxSOd Dec}), according to lemma \ref{decomposition lemma} the only contributing terms to the trivial representation are $\Delta^{1}\otimes\Delta^{1}$, $\Delta^{AS}\otimes\Delta^{AS}$ and  $\Delta^{S}\otimes\Delta^{S}$, each of which appears once. Therefore, the tensor $K_{ijkl}$ that is invariant under $\mathrm{SO}(d)$ belongs to the three dimensional IR subspace. We can form a basis in it by combining Kronecker delta tensors $\delta_{ij}$. There are three different ways to combine them into a four-fold tensor, therefore:
\begin{equation}\label{4tensor}
K_{ijkl}=\alpha \delta_{ij}\delta_{kl}+\beta \delta_{ik}\delta_{jl}+\gamma \delta_{il}\delta_{jk}.
\end{equation}

From the analysis given in the main text, only $d=3$ case exhibits non-trivial invariant dynamics. The most general dynamical law for the global state $\psi=(\bx,\by,T,\Lambda)$ is given by:
\begin{eqnarray}\label{d=3 dynamics1}
\frac{dx_i}{dt}&=&a\epsilon_{ijk}T_{jk}+L^{(1)}_{in}\lambda_n,\\\label{d=3 dynamics2}
\frac{dy_i}{dt}&=&b\epsilon_{ijk}T_{jk}+L^{(2)}_{in}\lambda_n,\\\label{d=3 dynamics3}
\frac{dT_{ij}}{dt}&=&-a\epsilon_{ijk}x_k-b\epsilon_{ijk}y_k+L^{(12)}_{ijn}\lambda_n+K_{ijkl}T_{kl},\\\label{d=3 dynamics4}
\frac{d\lambda_n}{dt}&=&Q_{nm}\lambda_m-L^{(1)}_{in}x_{i}-L^{(2)}_{in}y_{i}-L^{(12)}_{ijn}T_{ij}.
\end{eqnarray}
Note that the reversibility requires $K_{ijkl}=-K_{klij}$. If we apply this constraint to the equation \eqref{4tensor}, we obtain $K=0$.

 Next we will find the consistent values for the constants $a,b,L^{(1)}_{in},L^{(2)}_{in},L^{(12)}_{ijn}$ such that the solutions to the dynamical equations \eqref{d=3 dynamics1}--\eqref{d=3 dynamics4} above always lead to non-negative probabilities in Eq.~\eqref{prob rule}. We look at the simplest case where all the couplings to global parameters are zero $L^{(1)}_{in}=L^{(2)}_{in}=L^{(12)}_{ijn}=0$. If our initial state is a product state, than the global parameters remain zero during the evolution and we can safely neglect them from the analysis. In other words, the solution to the dynamical equations admits local tomography ($\Lambda=0$) and it can be found by solving the following set of equations:
\begin{eqnarray}\label{local dynamics}
\frac{dx_i}{dt}&=&a\epsilon_{ijk}T_{jk},\\
\frac{dy_i}{dt}&=&b\epsilon_{ijk}T_{jk},\\
\frac{dT_{ij}}{dt}&=&-a\epsilon_{ijk}x_k-b\epsilon_{ijk}y_k.
\end{eqnarray}

Let us find the solution for the initial conditions $\vec{\psi}^{\pm}(0)=\{\be_3,\pm\be_3,\pm\be_3\be_3^{\TT}\}$, where $\be_3=(0,0,1)^{\TT}$. The only components that evolve in time are $x_3(t)$, $y_3(t)$ and $T_{12}(t)=-T_{12}(t)$, hence the solution has the form:
\begin{equation}\label{pmsolution}
\psi^{\pm}(t)=\left\{\left(
                      \begin{array}{c}
                        0 \\
                        0 \\
                        x^{\pm}(t) \\
                      \end{array}
                    \right),\left(
                      \begin{array}{c}
                        0 \\
                        0 \\
                        y^{\pm}(t) \\
                      \end{array}
                    \right),\left(
                              \begin{array}{ccc}
                                0 & \tau(t) & 0 \\
                                -\tau(t) & 0 & 0 \\
                                0 & 0 & \pm1 \\
                              \end{array}
                            \right)
\right\},
\end{equation}
where $x^{\pm}(t),y^{\pm}(t)$ and $\tau(t)$ are the solutions to:
\begin{eqnarray}
\frac{dx^{\pm}}{dt}&=&2a\tau,\\
\frac{dy^{\pm}}{dt}&=&2b\tau,\\
\frac{d\tau}{dt}&=&-a x^{\pm}-b y^{\pm}.
\end{eqnarray}
Note that the state $\vec{\psi}^{\pm}(t)$ has to be physical state, that is, probability of equation \eqref{prob rule} is non-negative $P_{12}(\vec{\psi}|~\ba,\bb)\geq0$ for arbitrary choice of local measurements $\ba$ and $\bb$. If we set $\ba=\be_3$ and $\bb=-\be_3$, the positivity condition reads $\frac{1}{4}(x^{\pm}(t)-y^{\pm}(t))\geq0$. Similarly for $\ba=-\be_3$ and $\bb=\be_3$ we have $\frac{1}{4}(-x^{\pm}(t)+y^{\pm}(t))\geq0$. This is possible only if $x^{\pm}(t)=y^{\pm}(t)$.

In order to eliminate $\tau(t)$ from the dynamical equations we find the second derivatives in time of $x^{\pm}$ and $y^{\pm}$. We obtain:
\begin{eqnarray}
\frac{d^2x^{\pm}}{dt^2}&=&-2a^2x^{\pm}-2aby^{\pm},\\
\frac{d^2y^{\pm}}{dt^2}&=&-2abx_i-2b^2y^{\pm}.
\end{eqnarray}
This set of equation leads to the symmetric solution $x^{\pm}(t)=y^{\pm}(t)$ only if $a^2=b^2$ or equivalently $b=\pm a$. Note that $a=-b$ case brings new symmetry to the set of dynamical equations, the invariance under particle swap. If one requires such a symmetry, the case $a=b$ can be safely eliminated. However, we will use another argument that has been used in the work of Ref.~~[\onlinecite{DakicBrukner}]. We distinguish two cases, and label different solution as $\vec{\psi}_{\mathrm{MQM}}^{\pm}(t)$ and $\vec{\psi}_{\mathrm{QM}}^{\pm}(t)$, for $a=b$ and $a=-b$ respectively. The label QM and MQM stands for \emph{quantum mechanics} and \emph{mirror quantum mechanics} and the meaning of notation we explain shortly.

It is straightforward to evaluate the solution of dynamical equations:
\begin{widetext}
\begin{eqnarray}
&&\psi_{\mathrm{MQM}}^{+}(t)=\left\{\left(
                      \begin{array}{c}
                        0 \\
                        0 \\
                        \cos2at \\
                      \end{array}
                    \right),\left(
                      \begin{array}{c}
                        0 \\
                        0 \\
                        \cos2at \\
                      \end{array}
                    \right),\left(
                              \begin{array}{ccc}
                                0 & -\sin2at & 0 \\
                                \sin2at & 0 & 0 \\
                                0 & 0 & 1 \\
                              \end{array}
                            \right)
\right\},~~~~
\psi_{\mathrm{MQM}}^{-}(t)=\left\{\left(
                      \begin{array}{c}
                        0 \\
                        0 \\
                        1 \\
                      \end{array}
                    \right),\left(
                      \begin{array}{c}
                        0 \\
                        0 \\
                        -1 \\
                      \end{array}
                    \right),\left(
                              \begin{array}{ccc}
                                0 & 0 & 0 \\
                                0 & 0 & 0 \\
                                0 & 0 & -1 \\
                              \end{array}
                            \right)
\right\},\\
&&\psi_{\mathrm{QM}}^{-}(t)=\left\{\left(
                      \begin{array}{c}
                        0 \\
                        0 \\
                        \cos2at \\
                      \end{array}
                    \right),\left(
                      \begin{array}{c}
                        0 \\
                        0 \\
                        -\cos2at \\
                      \end{array}
                    \right),\left(
                              \begin{array}{ccc}
                                0 & \sin2at & 0 \\
                                -\sin2at & 0 & 0 \\
                                0 & 0 & -1 \\
                              \end{array}
                            \right)
\right\},~~~
\psi_{\mathrm{QM}}^{+}(t)=\left\{\left(
                      \begin{array}{c}
                       0 \\
                        0 \\
                        1 \\
                      \end{array}
                    \right),\left(
                      \begin{array}{c}
                        0 \\
                        0 \\
                        -1 \\
                      \end{array}
                    \right),\left(
                              \begin{array}{ccc}
                                0 & 0 & 0 \\
                                0 & 0 & 0 \\
                                0 & 0 & -1 \\
                              \end{array}
                            \right)
\right\}.
\end{eqnarray}
\end{widetext}

Our goal is to show that $\psi_{\mathrm{QM}}$ and the associate dynamics corresponds to quantum mechanics for two qubits, whereas $\psi_{\mathrm{MQM}}$ belongs to so called \emph{mirror quantum mechanics}~\cite{DakicBrukner}. The later case has the set of states obtained by partial transpose of two-qubit states. We introduce the matrix representation of $\vec{\psi}=(\bx,\by,T)$:
\begin{equation}
\rho(\vec{\psi})=\frac{1}{4}(\openone\otimes\openone+x_i\sigma_i\otimes\openone+y_i\openone\otimes\sigma_i+T_{ij}\sigma_i\otimes\sigma_j),
\end{equation}
where $\sigma_i,~i=1,2,3$ are the Pauli matrices. Straightforward calculation shows that $\rho(\psi_{\mathrm{QM}}^{-}(t))=\ket{\psi(t)}\bra{\psi(t)}$ is a density matrix, furthermore, it is a pure quantum state, where $\ket{\psi(t)}=\cos at\ket{0}\ket{1}+i\sin at\ket{1}\ket{0}$. Similarly, one can show that the matrix representation of mirror state $\psi_{\mathrm{MQM}}^{+}(t)$ is a non-quantum state (unless $\psi_{\mathrm{MQM}}^{+}(t)$ is product state) that can be obtained from $\psi_{\mathrm{QM}}^{-}(t)$ by applying total inversion $\by\mapsto-\by$ on the second spin. Note, that is a non-quantum operation. Mirror quantum mechanics is shown to be mathematically inconsistent theory for the tripartite case~\cite{DakicBrukner}. Therefore we will adopt only quantum solution.

The set of dynamical equations \eqref{local dynamics} has the corresponding matrix form:
\begin{equation}
\frac{d\rho(\vec{\psi})}{dt}=i[H_{12},\rho(\vec{\psi})],
\end{equation}
where $H_{12}$ is the Heisenberg spin-spin interaction $H_{12}=\frac{a}{2}\vec{\sigma_1}\vec{\sigma_2}=\frac{a}{2}\sum_{i=1}^{3}\sigma_i\otimes\sigma_i$.


\begin{thebibliography}{99}
\bibitem{footnote3} The argument for choosing to consider the generalized spins can be made more rigorous. In the spirit of operational theories we assume that every continuous reversible transformation (e.g. rotation) of macroscopic devices in the physical space generates a continuous reversible transformation of the system in the state space between two pure states. This excludes the ``box-world''~\cite{PRBox,Barrett} systems as they have discrete state spaces. The systems with relaxed uncertainty relations~\cite{Wehner} are also excluded since they require non-linear transformations.
\end{thebibliography}
\end{document}